\documentclass[manuscript]{acmart}

\makeatletter
\let\@authorsaddresses\@empty
\makeatother

\setcopyright{none}
\renewcommand\footnotetextcopyrightpermission[1]{}

\AtBeginDocument{%
  \providecommand\BibTeX{{%
    \normalfont B\kern-0.5em{\scshape i\kern-0.25em b}\kern-0.8em\TeX}}}

\acmBooktitle{ACM Transactions on Architecture and Code Optimization}
\acmPrice{15.00}
\acmISBN{978-1-4503-XXXX-X/18/06}

\newcommand{\ignore}[1]{}
\usepackage{fancyhdr}
\usepackage[normalem]{ulem}
\usepackage{url}
\usepackage{color}
\usepackage{soul}
\usepackage{wrapfig}

\usepackage{booktabs} 
\usepackage{comment}
\usepackage{subcaption}
\usepackage{balance}
\usepackage[english]{babel}
\usepackage[]{footmisc}

\usepackage{tikz}

\newcommand{\squishlistnum}{  
 \newcounter{qcounter}
 \begin{list}{\arabic{qcounter}:~}{\usecounter{qcounter}}
  { \setlength{\itemsep}{0pt}
     \setlength{\parsep}{0pt}
     \setlength{\topsep}{0pt}
     \setlength{\partopsep}{0pt}
     \setlength{\leftmargin}{1em}
     \setlength{\labelwidth}{1em}
     \setlength{\labelsep}{0.5em} } }

\newcommand{\squishlist}{ 
 \begin{list}{$\bullet$}  
  { \setlength{\itemsep}{2pt}
     \setlength{\parsep}{0pt}
     \setlength{\topsep}{3pt}
     \setlength{\partopsep}{0pt}
     \setlength{\leftmargin}{1em}
     \setlength{\labelwidth}{1em}
     \setlength{\labelsep}{0.5em} } }

\newcommand{\squishend}{
  \end{list}  }
  
\newcommand{\runinsec}[1]{\noindent\textbf{#1:} }

\ifdefined\isRelease

\newcommand\mehdi[1]{}
\newcommand\david[1]{}
\newcommand\rakesh[1]{}

\else

\newcommand\david[1]{\noindent{\color{red} {\bf \fbox{david}} {\it#1}}}
\newcommand\mehdi[1]{\noindent{\color{magenta} {\bf \fbox{mehdi}} {\it#1}}}
\newcommand\rakesh[1]{\noindent{\color{magenta} {\bf \fbox{rakesh}} {\it#1}}}

\fi

\begin{document}

\title{Freeway to Memory Level Parallelism in Slice-Out-of-Order Cores}
\titlenote{
    \textit{Extension of Conference Paper.} The article is an extension of our conference paper entitled “Freeway:Maximizing MLP for Slice-Out-of-Order Execution”, which was presented at HPCA-2019~\cite{freeway}. The key contributions of this article, over the conference version, are as follows:
  \squishlist
  \item Prior work evaluated slice-out-of-order (sOoO) cores only at small instruction windows. This work analyzes their scalability across the full spectrum of window sizes (i.e. from wimpy to brawny cores). We show that sOoO cores are less effective at large (wide/deep) instruction windows because MLP contributes less to the overall performance as the window size grows. This analysis, together with the fact that Freeway captures the bulk of available MLP, implies that the future research in scaling sOoO cores should focus on ILP rather than capturing the missed MLP opportunities. (Section~\ref{sec::WindowSizeSens})
   \item We assess the resource utilization and L1 cache latency tolerance of Freeway over the previous state-of-the-art, Load Slice Core (LSC). We show that Freeway requires one-sixth as many instruction queue entries and can tolerate 1.5x higher L1 latencies while providing same or better performance than LSC. These results imply that Freeway is a better choice than prior sOoO cores in resource constrained environments. (Section~\ref{sec::Qsizesens} and Section~\ref{sec::L1LatSens})
   \item We analyze the contribution of different sources of MLP i.e. Freeway, our proposed core design, and LLC prefetcher. Our results show that they mostly complement each other, although Freeway generates much more MLP than the LLC perfetcher. These results also imply that in area constrained designs it is better to invest area in Freeway rather than in an LLC prefetcher. (Section~\ref{sec::MLPBreakdown})
   \squishend
   * This work was done while Mehdi Alipour was at Uppsala University.
}

\author{Rakesh Kumar}
\affiliation{%
   \institution{Norwegian University of Science and Technology (NTNU)}
   \city{Trondheim}
   \state{Trondelag}
   \country{Norway}}
\email{rakesh.kumar@ntnu.no}

\author{Mehdi Alipour}

\affiliation{%
   \institution{Ericsson Research}
   \city{Lund}
   \state{Lund}
   \country{Sweden}}
\email{mehdi.alipour@ericsson.com}

\author{David Black-Schaffer}
\affiliation{%
   \institution{Uppsala University}
   \city{Uppsala}
   \state{Uppsala}
   \country{Sweden}}
\email{david.black-schaffer@it.uu.se}

\begin{abstract}

Exploiting memory level parallelism (MLP) is crucial to hide long memory and last level cache access latencies. While out-of-order (OoO) cores, and techniques building on them, are effective at exploiting MLP, they deliver poor energy efficiency due to their complex and energy-hungry hardware.
This work revisits slice-out-of-order (sOoO) cores as an energy efficient alternative for MLP exploitation. sOoO cores achieve energy efficiency by constructing and executing \textit{slices} of MLP generating instructions out-of-order only with respect to the rest of instructions; the slices and the remaining instructions, by themselves, execute in-order. However, we observe that existing sOoO cores miss significant MLP opportunities due to their dependence-oblivious in-order slice execution, which causes dependent slices to frequently block MLP generation.
To boost MLP generation, we introduce Freeway, a sOoO core based on a new dependence-aware slice execution policy that tracks dependent slices and keeps them from blocking subsequent independent slices and MLP extraction. The proposed core incurs minimal area and power overheads, yet approaches the MLP benefits of fully OoO cores. Our evaluation shows that Freeway delivers 12\% better performance than the state-of-the-art sOoO core and is within 7\% of the MLP limits of full OoO execution.

\end{abstract}




\maketitle

\section{Introduction}
\label{sec:intro}

Today's power-constrained systems face challenges in generating memory level parallelism (MLP) to hide the increasing access latencies across the memory hierarchy~\cite{memWall}. Historically, memory latency has been addressed through multilevel cache hierarchies to keep the frequently used data closer to the core. While cache hierarchies provide lower-latency in L1 caches, they have grown in complexity to the point where the 40-60 cycles it takes to access the last level cache has itself become a bottleneck. Therefore, exploiting MLP, by overlapping cache/memory accesses to hide the latency of later requests in the ``shadow'' of earlier requests, across the entire hierarchy is crucial for performance. However, the traditional approaches to extract MLP, such as out-of-order (OoO) or run-ahead execution, are not energy-efficient. 

The standard means of extracting MLP is out-of-order (OoO) execution, as it enables parallel memory accesses by executing independent memory instructions from anywhere in the issue window. However, the ability to identify, select, and execute independent instructions in an arbitrary order, while maintaining program semantics, requires complex and energy-hungry hardware structures. For example, one of the key enablers of OoO execution, the OoO issue queue, is typically built using content addressable memories (CAMs), whose power consumption grows super-linearly with queue depth and issue width.

State-of-the-art MLP extraction techniques aim to improve performance by increasing the amount of MLP extraction beyond the OoO execution. However, they fail to deliver energy efficiency primarily because they build upon already energy-hungry OoO execution and further introduce significant additional complexity of their own. For example, Runahead Execution~\cite{runahead}, which was originally proposed to improve the data cache performance in in-order cores~\cite{runaheadIO}, continues to extract MLP after an OoO core stalls, but requires additional resources for checkpointing and restoring states, tracking valid and invalid results, psuedo instruction retirement, and a runahead cache. This additional complexity entails a significant energy overhead.

To minimize the energy cost of MLP exploitation, a new class of cores, called \emph{slice-out-of-order} (sOoO) cores, builds on energy efficient in-order execution and adds just enough support for MLP extraction. These cores first construct groups, or \textit{slices}, of MLP generating instructions. A slice consists of one memory access instruction, i.e. load or store, and all the instructions required for its address generation. The slices and non-slice instructions are then dispatched to and scheduled from separate in-order instruction queues (IQs). As the instructions in one queue can bypass the instructions in the other queue, the slices and non-slice instructions execute out-of-order with respect to each other. However, by themselves, they still execute in-order as younger instructions cannot bypass the older instructions in the same in-order queue. Thus, by enabling MLP generating slices to bypass the rest of the potentially stalled instructions, sOoO cores extract significant MLP. Yet since they only support limited, coarser-grained out-of-order execution, they incur only a fraction of the energy cost of full, per-instruction out-of-order execution. 

The state-of-the-art sOoO core, the Load Slice Core (LSC)~\cite{LSC}, builds on an in-order stall-on-use core. LSC identifies MLP generating instructions using a small hardware table. To enable these instructions to execute out-of-order with respect to the rest of the instructions, LSC adds an additional in-order instruction queue, called the bypass queue (B-IQ). By restricting the out-of-order execution to choosing between the heads of two in-order instruction queues (the main, or A-IQ, and the bypass B-IQ), LSC minimizes energy requirements while still providing MLP.

Though sOoO cores are highly energy efficient, they fall noticeably behind OoO cores in terms of MLP extraction. Our key observation is that \textit{dependent-slices limit MLP extraction opportunities.} We define a dependent slice to be the one that contains at least one instruction that depends on the load instruction of another slice, called \emph{producer slice}. Dependent-slices limit MLP opportunities because, for example, when a dependent slice reaches the head of the in-order B-IQ in LSC, it blocks any further MLP generation by stalling the execution of subsequent, possibly independent, slices until the load instruction of its producer slice receives data from the memory hierarchy. Our analysis reveals that, in LSC, dependent slices block MLP generation for up to 83\% of the execution time (average 23\%). More importantly, the MLP loss is not just caused by the long stalling dependent slices whose producers miss in the on-chip caches. We demonstrate that, counter-intuitively, the dependent slices cause significant MLP loss even if they only stall for a few cycles: our results show that about 65\% of the dependent slice-induced MLP loss is caused by slices whose producers hit in the L1 cache. Together, these results demonstrate that dependent slices are a serious bottleneck in LSC.

This work addresses the fundamental limitation of the state-of-the-art sOoO core's ability to extract MLP: \textit{dependence-oblivious} first-in first-out (FIFO) slice execution causes dependent slices to delay the execution of subsequent independent slices. We propose to abandon the FIFO slice execution model in favor of a \textit{dependence-aware} slice scheduling model. Our proposed model tracks slice dependencies in hardware to identify dependent slices and steers them out of the way of the independent ones. As a result, the independent slices execute without stalling and expose more MLP.

To achieve this, we introduce Freeway, an energy efficient core design powered by a dependence-aware slice scheduling policy for boosting MLP and performance. Freeway tracks inter-slice dependencies with minimum additional hardware (one bit per entry in Register Dependence Table) to filter out the dependent slices. These slices are then steered to a new in-order queue, called the yielding queue (Y-IQ), where they wait until their producers finish execution. Such slice segregation clears the way for independent slices to generate more MLP as they no longer stall behind the dependent slices. Overall, Freeway delivers a substantial MLP boost by unblocking independent slice execution with minimal additional hardware resources. Our main contributions include:

\squishlist
\item Identifying that the dependence-oblivious FIFO slice execution is a major bottleneck to MLP generation in existing sOoO cores. We further demonstrate that dependent slices limit MLP even if they stall only for a few cycles (i.e. their producers hit in the L1 cache).
\item Proposing a new dependence-aware slice execution policy that executes independent slices unobstructed by tracking and keeping the dependent slices out of their way, hence boosting MLP.
\item Introducing the Freeway core design that employs minimal additional hardware to implement the dependence-aware slice execution: one bit per entry in Register Dependence Table, 7-bits per entry in Store Buffer, a FIFO instruction queue, and some combinational logic.
\item Demonstrating that Freeway provides 12\% more performance than the state-of-the-art sOoO core and is within 7\% of the MLP limits of full OoO execution. We also analyze the remaining bottlenecks that cause this 7\% performance gap and show that mitigating them brings minimal performance returns on resource investment.
\item Showing better resource utilization and latency tolerance for Freeway than LSC. Freeway needs six times fewer IQ entries and tolerates 1.5x higher L1 cache latency while providing better or same performance as LSC.
\item Analyzing the scalability of sOoO cores over the full spectrum of instruction window sizes from wimpy to brawny cores. We show that sOoO cores are less effective at large (wide/deep) instruction windows because the MLP contributes less to overall performance as the window size grows. This result, along with the fact that Freeway captures the bulk of MLP opportunity, implies that the future research in scaling sOoO cores should focus on ILP rather than capturing the missed MLP opportunities.

\squishend

\section{Background and Motivation}
\label{sec:background}

\subsection{MLP vs Energy: InO, OoO, and slice-OoO cores}

Existing core designs force a trade-off between MLP and energy efficiency. For example, an in-order (InO) core can be highly energy efficient, but it is unable to generate significant MLP, and therefore delivers poor performance. In contrast, OoO cores are generally good at extracting MLP, but at the cost of (much) lower energy efficiency. To exploit MLP while delivering high energy efficiency, a recent design, the Load Slice Core (LSC)~\cite{LSC}, proposed a new approach of \textit{slice}-out-of-order (sOoO) execution. LSC builds on an efficient in-order core and employs separate instruction queues, AIQ and B-IQ, for non-MLP and MLP generating instructions, respectively. This enables MLP generating instructions in the B-IQ to bypass the potentially stalled load consumers in the A-IQ. By exposing MLP in this way, LSC avoids much of the complexity of full OoO architectures.

\begin{figure*}[t]
    \centering
    \includegraphics[width=\textwidth, trim=10 60 30 140, clip]{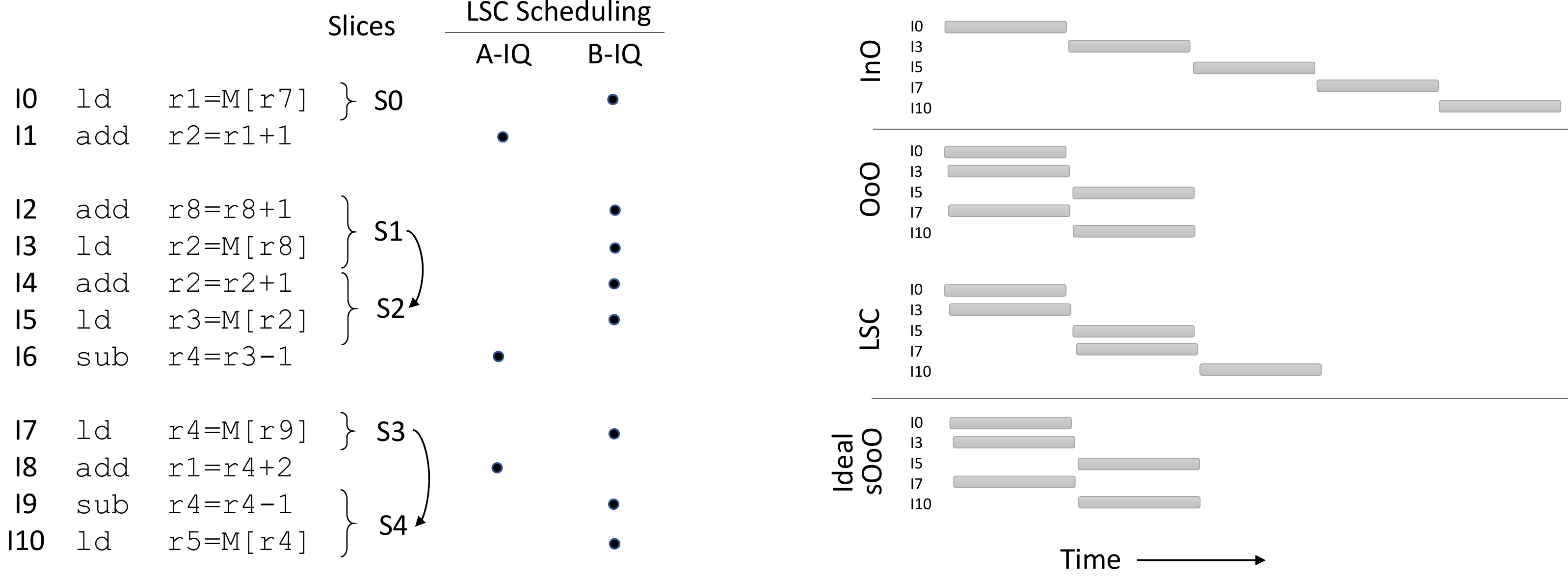}
    \vspace{-0.1in}
    \caption{Overlapping memory accesses in InO, OoO, LSC, and Ideal sOoO. The arrows show inter-slice dependencies.}
    \label{fig:MLPCartoon}
    \vspace{-0.2in}
\end{figure*}

\runinsec{MLP Extraction}Figure~\ref{fig:MLPCartoon} shows how the sOoO execution of LSC fairs against InO and OoO cores in exploiting MLP. As a stall-on-use InO core stalls on the first use of the value being loaded from memory, it serializes all the loads in this example, resulting in no MLP. The OoO core is able to extract the maximum MLP by overlapping the execution of independent load instructions (I0, I3 and I7). When these loads' data returns, their dependent loads (I5 and I10) are also overlapped. The sOoO execution of LSC falls between the InO and OoO cores. LSC overlaps the execution of the first two load instructions (I0 and I3) as the B-IQ enables I2 and I3 to bypass the stalled instructions (I1) in the A-IQ.

Figure~\ref{fig:MLPCartoon} also demonstrates a major limitation of LSC: it is effective in extracting MLP only when the \textit{slices are independent}. A dependent slice at the head of the B-IQ stalls MLP extraction by blocking the execution of the subsequent independent slices. In Figure~\ref{fig:MLPCartoon}, slice S2, a dependent slice as it depends on the load in S1, stalls the B-IQ and delays the execution of the next independent slice S3 until its producer slice S1 receives data from the memory hierarchy. Such slice dependencies limit MLP and overall performance. However, an Ideal sOoO core, that allows fully out-of-order execution among slices, would eliminate this limitation. As shown in Figure~\ref{fig:MLPCartoon}, an ideal sOoO core matches the MLP generation of a full OoO core.

\runinsec{Energy Consumption}LSC's sOoO execution is implemented using simple hardware components: FIFO queues and small tables for tracking slices. As a result, it only slightly increases the area and power consumption compared to an already small InO core. In contrast, the energy requirements of OoO cores are substantially higher due to the use of complex structures, such as CAMs. Indeed, previous research~\cite{complexityEffective} has shown that the ability to select arbitrary instructions from IQ is one of the most energy consuming tasks in OoO cores. Carlson et.~al.~\cite{LSC} concluded that LSC incurs only 15\% area and 22\% power overheads over an InO core (ARM Cortex-A7), whereas an out-of-order core (ARM Cortex-A9) requires 2.5x area and 12.5x power compared to the same in-order core.

The slice-out-of-order execution in LSC is a promising step towards energy efficient MLP extraction. However, LSC's strict FIFO execution of slices limits its potential to extract MLP in the case of dependent slices. To understand this limitation, we next explore its impact on performance.

\subsection{Potential for MLP extraction}
To quantify the potential MLP available in a sOoO core, we compare LSC, with its in-order B-IQ, to a LSC with a fully out-of-order B-IQ (Ideal-sOoO). While an out-of-order B-IQ would be impractical (it would defeat the efficiency goal of avoiding the complexity of out-of-order instruction selection), it allows us to observe the maximum MLP gains possible if the independent slices can bypass other stalled slices.
(Our simulation methodology, including microarchitectural parameters, is detailed in Section~\ref{sec:method}.)

\begin{figure}[t]
    \centering
    \includegraphics[width=0.8\textwidth, trim=200 135 190 150, clip]{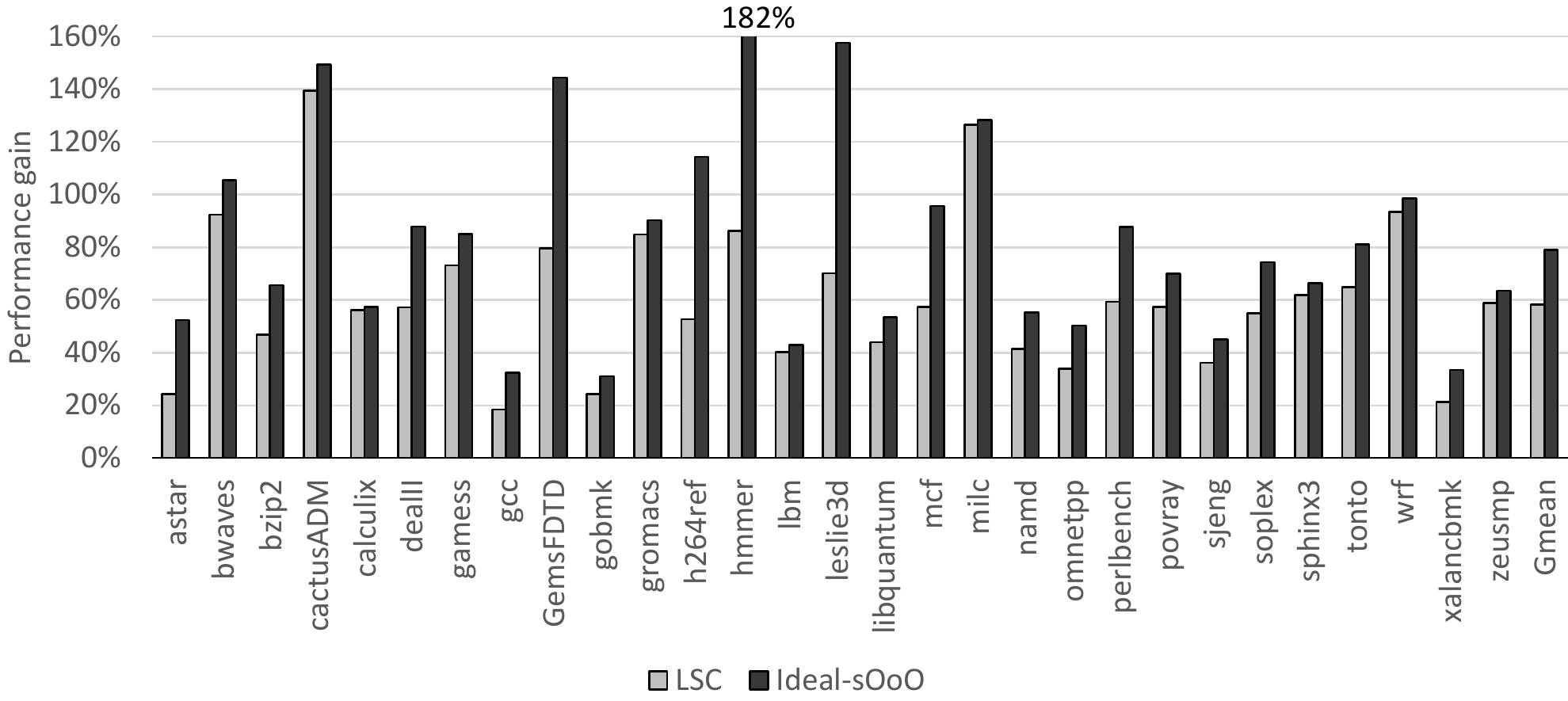}
    \vspace{-0.15in}
    \caption{Performance gain for LSC and Ideal-sOoO over InO execution.} 
    \label{fig:opportunity}
    \vspace{-0.15in}
\end{figure}

Figure~\ref{fig:opportunity} shows the performance gains obtained by LSC and Ideal-sOoO (LSC with a fully out-of-order B-IQ) over an InO core. The relative difference between the two shows the opportunity missed by LSC due to its FIFO slice execution.
As the figure shows, the performance difference between Ideal-sOoO and LSC is 20\%, geometric mean, and more than 50\% on \texttt{GemsFDTD}, \texttt{h264ref}, \texttt{hmmer}, and \texttt{leslie3d}, due to relatively larger numbers of dependent slices. However, there is little gain for \texttt{calculix}, \texttt{lbm}, and \texttt{milc}, as they have fewer dependent slices (See Section~\ref{sec::stallDist}). Overall, the majority of the workloads demonstrate considerable performance opportunity if we can eliminate the dependent slice bottleneck.

\subsection{Sources of stalls in the bypass queue}
\label{sec::stallDist}

For a deeper understanding of the microarchitectural bottlenecks limiting MLP extraction in LSC, we examine the stall sources afflicting the B-IQ and categorize them as follows:

\squishlist
\item \textbf{Slice Dependence Stalls:} A dependent slice at the B-IQ head is waiting for its producer to receive data from the memory hierarchy. 
\item \textbf{Empty B-IQ Stalls:} There are no instructions (slices) in the B-IQ.
\item \textbf{Load-store Aliasing Stalls:} A load at B-IQ head cannot be issued because an older store in the A-IQ is waiting to write to the same address (true alias). This might happen because, in LSC, store data calculations and the store operations themselves go to the A-IQ, whereas, the store address calculation goes to the B-IQ.
\item \textbf{Other Stalls:} Intra-slice dependencies, unresolved store addresses blocking younger loads, etc.
\squishend

For this study, we assume an ideal core front-end (no instruction cache or BTB misses and a perfect branch predictor) to isolate the slice execution bottlenecks. 

\begin{figure}[t]
    \centering
    \includegraphics[width=0.8\textwidth, trim=65 170 60 175, clip]{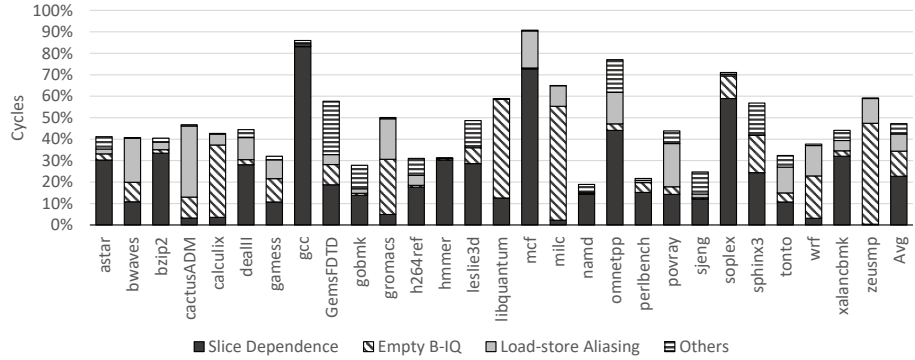}
    \vspace{-0.15in}
    \caption{Percentage of execution time the issue stage is stalled in LSC, and the breakdown of stall sources.}
    \label{fig:stallDist}
    \vspace{-0.15in}
\end{figure}

Figure~\ref{fig:stallDist} shows the breakdown of stall cycles (when no instruction is issued from either the A-IQ or the B-IQ) as a fraction of overall execution time. The figure reveals that instruction issue is stalled for 47\% of the execution time on average, and \textit{Slice Dependence Stalls} are responsible for almost half of these stalls. The \textit{Slice Dependence Stalls} are particularly significant in \texttt{gcc}, \texttt{mcf}, \texttt{soplex}, and \texttt{hmmer}, where they account for more than 80\% of all stalls. Notice that \texttt{gcc} and \texttt{mcf} are the most severely affected workloads, yet they are not the ones that show the highest performance opportunity with Ideal sOoO execution (Figure~\ref{fig:opportunity}). The reason is that the performance opportunity is a function of not only the number of stalls caused by dependent slices, but also where their producer slices hit in the memory hierarchy. As shown in Figure~\ref{fig:sliceDepStalls}, the majority of producer slices, in \texttt{gcc} and \texttt{mcf}, miss in the on-chip cache hierarchy and must be loaded from memory. This long memory latency stalls instruction retirement, and therefore causes the instruction window to fill which blocks further MLP generation and limits performance. In this work, we use the term instruction window to refer to the window of all in-flight instructions, i.e., the instructions that have been dispatched but not yet committed. Scoreboard and reorder buffer (ROB) are the typical hardware implementations of instruction window in sOoO and OoO cores, respectively.

\textit{Empty B-IQ Stalls} are the second largest source of stalls. We observe that the primary reason for the B-IQ to be empty is a full instruction window and the oldest instruction is not ready to retire. As a result, no new instructions can enter either instruction queue. This could be remedied through larger instruction windows or methods such as Runahead Execution~\cite{runahead}. The third largest source of stalls are \textit{Load-store Aliasing Stalls}, and they are particularly severe in \texttt{bwaves}, \texttt{cactusADM}, \texttt{gromacs}, \texttt{lbm}, and \texttt{povray}. However, they account for only 8\% of execution time. 
 
Looking at the potential of an Ideal-sOoO, we see that LSC misses about 20\% performance opportunity. The majority of this loss is due to Slice Dependence Stalls. Next, we analyze memory slice behaviour to mitigate this bottleneck.

\section{Addressing Slice Dependence}
\label{sec:unlockMLP}

A generic approach to handle dependent slices is to get them out of the way of the independent slices by buffering them outside of the B-IQ. To this end, we next study the slice behaviour to understand which dependent slices should be buffered and where they should be buffered.

\subsection{Which dependent slices to buffer?}
\label{sec:WhatToBuffer}

Intuitively, only the dependent slices that stall the B-IQ for many cycles need to be buffered. Such long stalls are typically due to slice's producers hitting in the LLC or memory. However, unintuitively, we found that 65\% of the \textit{Slice Dependence Stalls} are caused by dependent slices that stall only for a few cycles as their producers \textit{hit} in the L1 cache. This demonstrates that even the relatively short L1 hit latency (4 cycles in our simulation) can significantly limit the MLP and performance in a sOoO core with strict FIFO slice execution. 

\begin{figure}[t]
    \centering
    \includegraphics[width=0.8\textwidth, trim=60 180 60 170, clip]{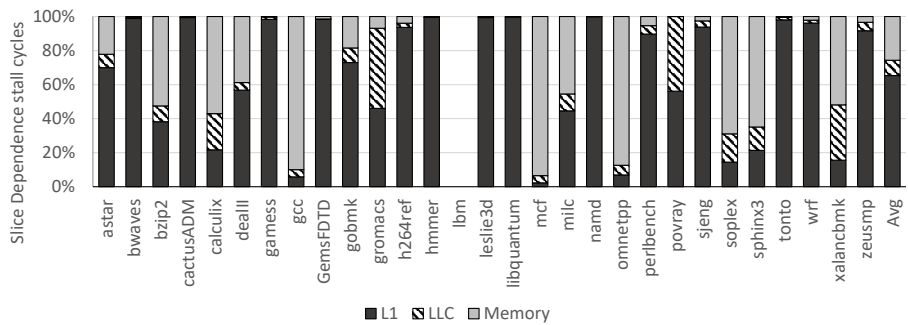}
    \vspace{-0.15in}
    \caption{Breakdown of \textit{Slice Dependence Stall} cycles based on the producer slice hit site in the memory hierarchy. \texttt{lbm} does not have any dependent slices.}\label{fig:sliceDepStalls}
    \vspace{-0.15in}
\end{figure}

Figure~\ref{fig:sliceDepStalls} shows the breakdown of \textit{Slice Dependence Stall} cycles based on the producer slice hit site in the memory hierarchy.
The results are especially interesting for workloads such as \texttt{hmmer}, where producer slices almost always hit in the L1 cache, and yet dependent slices are responsible for more than 96\% of all stall cycles, which accounts for about 31\% of the execution time (Figure~\ref{fig:stallDist}).

These results suggest that it is important to buffer all dependent slices, even those that only stall for the duration of an L1 hit. Interestingly, this also suggests that in many cases we only need to buffer the dependent slices for a few cycles (to cover L1 latency) to achieve much of the MLP benefit. If such limited buffering is sufficient, it would suggest we can achieve these benefits at a low implementation cost.

\subsection{Where to buffer?}
\label{sec::whereToBuffer}
To mitigate the slice dependence bottleneck, the dependent slices need to be kept in a separate buffer to prevent them from stalling the B-IQ. However, traditional instruction buffers, such as the Waiting Instruction Buffer~\cite{WIB}, are complex, energy intensive, and are designed to buffer instructions for longer time intervals, such as during LLC misses. In addition, those designs require the instructions to be inserted back to the main IQ before issuing them for execution~\cite{WIB, LTP}. The extra energy and latency of re-inserting instructions is particularly costly for instruction slices which will only be buffered for a few cycles.

\begin{wrapfigure}{R}{0.65\textwidth}
\vspace{-0.2in}
    \centering
    \includegraphics[width=0.65\columnwidth, trim=30 220 270 140, clip]{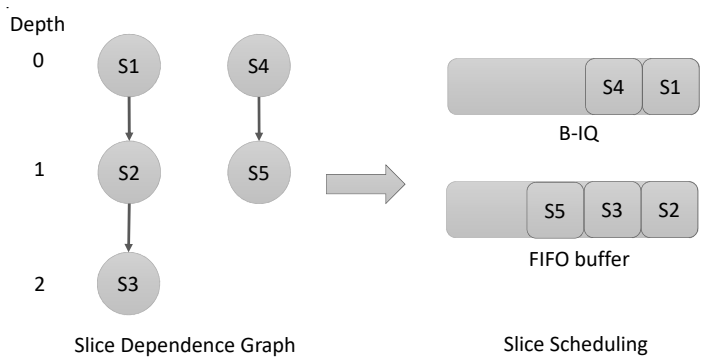}
    \vspace{-0.4in}
    \caption{Slice dependence graph with dependence depth (left) and slice scheduling to different queues (right). S1 - S5 are slices.}\label{fig:depSliceCartoon}
    \vspace{-0.2in}
\end{wrapfigure}

A simple FIFO queue is an attractive alternative instruction buffer due to its low complexity and energy cost. However, as instructions can only be picked from the head of the FIFO queue, buffering all dependent slices in a single queue might cause a bottleneck if the younger slices become ready for execution before the older slices. This occurs for primarily two reasons: First, if a younger slice has fewer slices before it in its dependence chain than a slice in an older chain. Or, second, if the producer of a younger slice hits closer to the core in the memory hierarchy than the producer of an older slice. To understand the implications of these effects, we analyze potential stall sources to determine if a single, cheap, FIFO queue is appropriate for buffering dependent slices.

\runinsec{Slice dependence depth} 
\label{sec:sliceDep}
Slices further down their dependence chains can potentially stall the execution of slices in a younger chain. To better understand this, we define the \textit{dependence depth} of a slice as the number of slices in the dependent slice chain leading up to it. For example, in Figure~\ref{fig:depSliceCartoon}, S1 and S4 are independent slices and start the dependent slice chain, hence their dependence depth is 0. Next, S2 and S5 are at dependence depth 1 as they have one slice ahead of them, S1 and S4, respectively. 

Using this definition, we observe that a younger slice with a lower dependence depth is likely to become ready before an older slice with higher dependence depth. For example, in Figure~\ref{fig:depSliceCartoon}, S3 (depth 2) will be ready only after both S2 and S1 have received their data, whereas S5 (depth 1) needs to wait only for S4. If all the slices hit at the same level in memory hierarchy, leading to similar execution times, S5, the younger slice, will be ready for execution before S3. However, it will be stalled behind S3 in the FIFO queue, thereby limiting MLP extraction. 

\begin{figure}[t]
    \centering
    \includegraphics[width=0.8\textwidth, trim=60 180 60 180, clip]{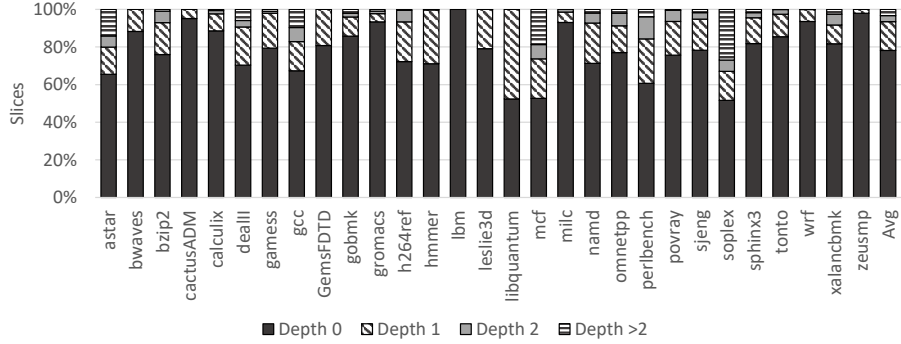}
    \vspace{-0.1in}
    \caption{Slice dependence depth distribution.}\label{fig:sliceDepth}
    \vspace{-0.1in}
\end{figure}

To understand the potential bottleneck due to such stalls, we study the slice dependence depth in our workloads in Figure~\ref{fig:sliceDepth}. As the figure shows, 78\% of all slices are independent slices (depth 0) and do not need to be buffered. Of the remaining slices that do need to be buffered, more than 72\% are at dependence depth 1. Therefore, as the majority of dependent slices are at the smallest depth of 1, the stalls caused by slices at larger dependence depths (6\% of all slices) are likely to be minimal.

\vspace{.02in}

\runinsec{Producer slice hit site} Even if dependent slices are at the same dependence depth, a younger slice can still become ready earlier than an older slice if its producer hits closer to the core in the memory hierarchy than the producer of the older slice. For the example in Figure~\ref{fig:depSliceCartoon}, S2 and S5 both are at dependence depth 1, but S5 may become ready earlier if its producer S4 hits in L1 and S2's producer S1 hits in LLC or farther. In this scenario, S5 will be stalled as S2 is blocking the head of the FIFO queue. 

\begin{figure}[t]
    \centering
    \includegraphics[width=0.8\textwidth, trim=60 180 60 170, clip]{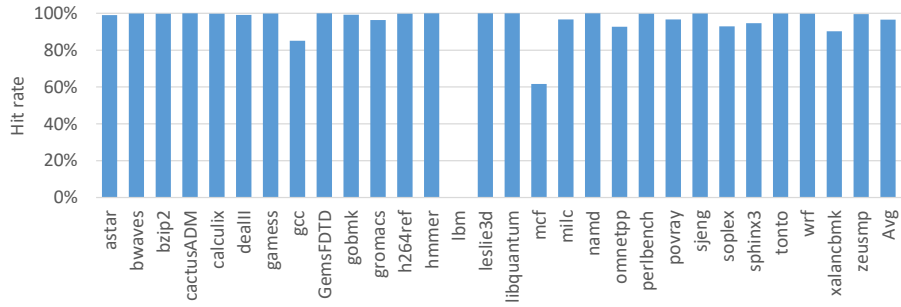}
    \vspace{-0.25in}
    \caption{L1 cache hit rate for producer slices at dependence depth 0. \texttt{lbm} does not have any dependent or producer slices.}\label{fig:prodSliceHitSite}
    \vspace{-0.15in}
\end{figure}

To understand the extent of the potential bottleneck, we study the hit site of producer slices with at least one dependent slice. For this study, we only consider the producer slices at dependence depth 0 because the majority of dependent slices are at depth 1 (i.e. dependence chain lengths of 2 slices).
Figure~\ref{fig:prodSliceHitSite} shows that more than 96\% of these producer slices hit in the L1 cache. Therefore, as the majority of producer slices hit at the same level, L1, the dependent slices are likely to become ready in the program order, and, hence, incur minimal stalls due to ready younger slices waiting behind stalled older ones.

\vspace{.02in}

Overall, Figures~\ref{fig:sliceDepth} and \ref{fig:prodSliceHitSite} suggest that a single FIFO queue for all dependent slices is sufficient to expose most of the potential MLP available from out-of-order slice execution. This is because most of the dependence slices are at the dependence depth one and the majority of producer slices hit in L1, indicating that it is unlikely that dependence slices will stall behind each other. (Results in Section~\ref{sec:anaRemOpp} validate that additional queues bring only minimal performance gains.)

\section{Freeway}
\label{sec:freeway}

Freeway is a new slice-out-of-order (sOoO) core designed to achieve the MLP benefits of full out-or-order execution. Freeway goes beyond prior work by addressing the sources of sOoO stalls identified in our analysis (Section~\ref{sec:unlockMLP}) to execute the majority of slices without stalling on dependent slices. Freeway requires only small changes over the baseline sOoO design (LSC), and thereby retains its low complexity. As a result, Freeway is able to substantially increase the exposed MLP and performance, while retaining a simple, and energy efficient design.

\begin{figure*}[t]
    \centering
    \includegraphics[width=0.85\textwidth, trim=35 220 10 140, clip]{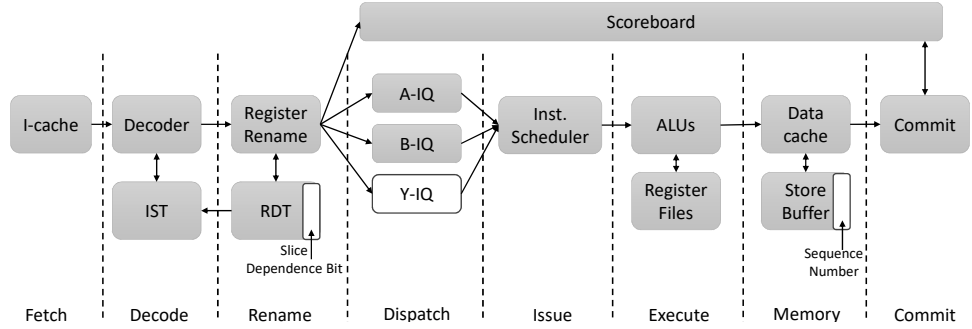}
    \caption{Freeway microarchitecture. New components are in white; Instruction scheduler issues from Y-IQ as well.}
    \label{fig:freeway}
    \vspace{-0.15in}
\end{figure*}

An overview of the Freeway microarchitecture is presented in Figure~\ref{fig:freeway}. The components common to both LSC and Freeway, such as the B-IQ, are shown in light gray. The additions required for Freeway are in white. As Freeway builds upon LSC, we first describe the baseline LSC microarchitecture before providing an overview of the Freeway design and, finally, a detailed discussion of the key design issues. 

\subsection{Baseline sOoO}
\label{sec:lsc}

The first sOoO design, the Load Slice Core, builds upon an energy efficient in-order, stall-on-use core. LSC identifies MLP generating instructions (slices) in hardware and executes them early (via the B-IQ) with minimum additional resources, thereby achieving both MLP and energy efficiency.

\runinsec{Slice construction} To identify MLP generating instructions efficiently, LSC leverages applications' loop behaviour to construct memory slices using iterative backward dependency analysis (IDBA)~\cite{LSC}, starting with the memory access instructions. In each loop iteration, the producers of the instructions identified in the previous iteration are added to the slice. LSC identifies the producers of an instruction through the Register Dependence Table (RDT), which maps each physical register to the instruction that last wrote to it. LSC then uses a simple PC-indexed Instruction Slice Table (IST) to track the instructions in slices. By building up slices via simple RDT look-ups over multiple loop iterations, LSC avoids the complexity and energy overheads of explicit slice generation techniques ~\cite{dynamicSlicePrecomp, sliceProc}.

Figure~\ref{fig:MLPCartoon} shows an example of slice construction. As soon as the decoder detects a load or store instruction, it starts slice construction. The construction of slice S1 in this figure starts with the load instruction I3. Once the load is detected, IBDA consults RDT to find its producers. In this case, RDT will report I2 to be a producer as it wrote to the input register, r8, of the load instruction I3. Further, I2 will be inserted into the IST as it has been identified to be a part of the slice. During the next iteration, I2 will hit in IST and IBDA will consult RDT to find its producers. However, I2 does not have any producers in the instruction window as its operands are already available when it is decoded. Therefore, IBDA will stop backwards traversal implying completion of slice construction.  Notice that a slice includes only data flow but not the control flow required to generate the address for memory access instruction.

\runinsec{Slice Execution} To exploit MLP, LSC adds an additional in-order \textit{bypass queue (B-IQ)} to the baseline InO core. The MLP generating instructions (slices), as identified by IST, are dispatched to the B-IQ, which enables them to bypass the instruction in the main instruction queue (A-IQ). For store instructions, only the address calculation is dispatched to the B-IQ so that their addresses are available earlier, and subsequent loads can be disambiguated. The data calculation and the store operation itself proceed to the regular instruction queue (A-IQ), as they rarely limit MLP. By allowing such limited out-of-order execution, MLP generating instructions can bypass (via the B-IQ) stalled instructions in the main instruction flow (the A-IQ). As a result, the sOoO execution of the LSC allows it to extract considerable MLP, without the energy cost of the full out-of-order execution.

\subsection{Freeway: Overview}

While LSC is effective in exploiting MLP when the memory slices are independent, dependent slices cause a serious bottleneck due to LSC's strict FIFO slice execution. As LSC mixes dependent slices with the independent ones in the B-IQ, it limits MLP by delaying the execution of independent slices stalled behind the dependent ones. To increase MLP, Freeway abandons LSC's FIFO slice execution and allows independent slices to execute out-of-order with respect to dependent ones with minimum additional hardware.

To enable out-of-order execution among slices, Freeway \textit{tracks} slice dependencies in hardware and \textit{separates} dependent slices from independent ones. Freeway uses the slice dependency information to \textit{accelerate} independent slice execution by reserving the B-IQ exclusively for them. To handle the dependent slices, Freeway introduces a new FIFO instruction queue called the \textit{yielding queue} (Y-IQ).  Dependent slices can then wait in the Y-IQ, yielding execution to the independent slices, until they become ready for execution. As our analysis in Section~\ref{sec:unlockMLP} demonstrated, the dependent slices mostly become ready in program order, therefore, ready dependent slices should rarely stall behind the non-ready ones. This characteristic allows us to use simple hardware to boost MLP by executing the majority of slices from both the B-IQ and Y-IQ without any stalls.

Freeway requires only minimal additional hardware, as shown in Figure~\ref{fig:freeway}, to support out-of-order slice execution: extending the RDT entries with one bit to track whether an instruction belongs to a dependent slice; adding a FIFO instruction queue (Y-IQ) for dependent slices; extending each store buffer entry with 7 bits and comparators to maintain memory ordering; and adding logic to issue instructions from the Y-IQ in addition to from the A-IQ and B-IQ.

\subsection{Freeway: Details}
We first describe the mechanism for tracking slice dependence and then provide details of instruction flow through Freeway, before discussing memory ordering requirements.

\subsubsection{Tracking dependent slices}
\label{sec:trackDepSlice}
We classify a memory slice as a dependent slice if it contains at least one instruction that depends on the \textit{load instruction} of an older slice. However, a slice is not classified as dependent if it depends on a non-load instruction of a older slice.
Freeway detects dependent slices in the \textit{Register Rename} stage by leveraging the existing data dependence analysis of the baseline core. These analyses are required by LSC to identify the instructions belonging to memory slices. The first instruction of a dependent slice can be identified trivially using the data dependence analysis as it is the instruction that receives at least one of its operands from a load instruction. Identifying the remainder of the dependent slice instructions is more involved as they may not be directly dependent on the load. Therefore, the dependence information must be propagated from the first dependent instruction to the memory access instruction terminating the slice.

Freeway extends LSC's RDT with a \textit{slice dependence bit} to propagate the dependence information through a slice. The dependence bit indicates whether the instruction reading a register would belong to a dependent slice or not. 
Initially the slice dependence bits are 0 for all RDT entries. When Freeway detects a load instruction, it sets the slice dependence bit of its destination register's RDT entry to 1, as any slice instruction reading this register would belong to a dependent slice. Subsequently, if the slice dependence bit for any of the source registers of an instruction is found to be 1, the instruction is marked as a dependent slice instruction. In addition, the dependent slice bit of its destination register's RDT entry is set to 1. This propagates the dependence information through the slice. As such, Freeway only requires 1 additional bit per RDT entry to identify the chain of dependent instructions constituting a dependent slice.

\subsubsection{Instruction Flow Through Freeway}
\runinsec{Front-end} The Freeway front-end is very simliar to that of LSC, but with the addition of dependent slice identification and tracking. As with LSC, after instruction fetch and pre-decode, the IST is accessed with the instruction pointer to check if an instruction belongs to a memory slice or not. This information is propagated down the pipeline to assist instruction dispatch. Next, register renaming identifies true data dependencies among instructions so that dependent instructions wait until their producers finish execution. 
At this point Freeway consults the RDT to determine if a memory slice instruction also belongs to a dependent slice, and passes on this information to the dispatch stage.

\runinsec{Instruction Dispatch} Freeway dispatches an instruction to one of the three FIFO instruction queues (A-IQ, B-IQ, or Y-IQ) based on the slice and dependence information received from the IST and RDT.
Loads, stores, and their address generating instructions, as identified by the IST are dispatched to the B-IQ if they belong to \textit{independent} memory slices. In contrast, if the RDT classifies them as part of a dependent slice, they are dispatched to the Y-IQ, where they wait until their producer slices finish execution. The rest of the instructions are dispatched to the A-IQ.
For Stores, as with LSC, the data calculation and the store operation itself are dispatched to the A-IQ. Whereas the address calculation goes to either the B-IQ or Y-IQ, based on its dependence status. Such split dispatching for stores enables their addresses to be available early so that the subsequent loads can be disambiguated against them and continue execution, if they access non-overlapping memory locations.

\runinsec{Back-end} The Freeway back-end selects instructions for execution from its three IQs (i.e., A-IQ, B-IQ, and Y-IQ), executes them, and ensures that they update architectural state in program order. As Freeway employs FIFO instruction queues, only the instructions at their heads can be scheduled for execution. If multiple IQs have ready instructions at their heads, instructions are scheduled using an age-based policy, i.e., older instructions are prioritized over the younger ones. We also studied prioritizing slices over non-slice instructions, however performance was similar to the age-based policy. Further, in each cycle, not all scheduled instructions need to come from the same IQ, rather they can come from any combination of IQs to fill the issue width.

Though Freeway employs only FIFO IQs for instruction scheduling, having multiple of them enables younger instructions in one IQ to bypass the older instructions stalled in the other IQs. As a result, instructions can be scheduled and executed out of program order. Therefore, Freeway needs to track instruction order to ensure that the instructions update the architectural state in program order. Like LSC, Freeway employs a Scoreboard to track instruction order. As shown in Figure~\ref{fig:freeway}, instruction are inserted into Scoreboard in program order at dispatch stage. As instructions finish their execution, which might be out of program order, they do not update the architectural state immediately, rather they record their completion in the Scoreboard. Only once an instruction reaches the head of scoreboard, it updates the architectural state and is taken off the Scoreboard. This ensures that the architectural state is always updated in program order even though instructions can execute out of order. To track a sufficient number of instructions, Freeway and LSC increase the size of Scoreboard over what is typical in an in-order core.

\vspace{-0.15in}

\subsubsection{Memory ordering}  
\label{sec:mem_order}
Before describing Freeway's mechanism to maintain memory ordering, we first discuss how the baseline LSC maintains this order. LSC computes memory addresses strictly in program order as all address calculations are performed via the FIFO B-IQ. Despite the FIFO address generation, younger loads can still bypass the older stores that are waiting in the A-IQ (recall that only the address calculation for stores is performed via B-IQ, whereas the store operation itself passes through the A-IQ). Therefore to avoid loads from bypassing the aliased stores, LSC incorporates a \textit{store buffer}. It inserts store addresses in to the store buffer so that they can be used to disambiguate the subsequent loads. LSC then issues loads to memory only if their address does not match any store address in the store buffer, thereby ensuring memory ordering.

This mechanism cannot be directly ported to Freeway to maintain memory ordering. This is because the strict FIFO address generation in LSC guarantees that all previous outstanding stores have their addresses in the store buffer when a load is about to be issued. Freeway, in contrast, allows independent memory slices to bypass the dependent ones waiting in the Y-IQ. As a result, a load may not check against an older store whose address calculation is still waiting in the Y-IQ and the address has not yet been written to the store buffer. To avoid this scenario, Freeway marks all loads and stores with a sequence number in program order. In addition, stores are allocated an entry in the store buffer at dispatch and the entry is later updated with the store address when available. As a result, loads that are about to be issued can look in the store buffer to check if all previous stores have computed their addresses. They only proceed to execution if there are no unresolved and aliasing stores. This simple store buffer extension maintains memory ordering while only requiring the addition of a small (depending on instruction window size) sequence number to the store buffer entries.

It is worth noting that, as with LSC, Freeway issues stores to the memory only when they are the oldest instruction in the instruction window. Therefore, such stores do not violate memory ordering even though they can bypass older loads waiting in the Y-IQ that access the same memory location. When such a bypassed load becomes ready, it checks the store buffer and finds a store with the same memory address. However, instead of forwarding data from the store, the load is issued to memory as the store is younger.

\vspace{-0.15in}
\section{Methodology}
\label{sec:method}

\begin{wraptable}{R}{0.5\textwidth}
\vspace{-0.2in}
	\centering 
	{\small
		\scalebox{0.8} {		
		\begin{tabular}{|c|c|}
			\hline  
			\hline  
			Core       & 
            \begin{tabular}{@{}c@{}}
			2GHz, 2-wide issue, 
			64-entry instruction window
            \end{tabular} \\			

			\hline  			
            Branch Predictor& Intel Pentium M-style~\cite{pentM} \\
			\hline  			
            Branch Penalty& 9 cycles (7 cycles for in-order core) \\
			\hline
			Functional Units& 2 Int, 1 VPU, 1 branch, 2 Ld/St (1+1) \\
			\hline
			L1-I & 
            \begin{tabular}{@{}c@{}}
            32 KB, 4-way LRU
            \end{tabular} \\
			\hline
			L1-D & 
            \begin{tabular}{@{}c@{}}
            32 KB, 8-way LRU, 4 cycle, 8 MSHRs
            \end{tabular} \\
			\hline
			LLC & 
			\begin{tabular}{@{}c@{}}
			512KB per core, 16-way LRU, \\
			avg 30-cycle round-trip latency
			\end{tabular}\\
			\hline
			 LLC Prefetcher  & stride-based, 16 independent streams\\
			\hline
			Main Memory & 4 GB/s, 45 ns access latency\\
			\hline
			\hline  			
		\end{tabular}
	  }
	}
	\caption{Microarchitectural parameters}
	\label{microParam}
	\vspace{-0.1in}
\end{wraptable}

To evaluate Freeway, we use the Sniper~\cite{sniper} simulator configured with a cycle-accurate core model~\cite{sniperCoreModels}. Sniper works by extending Intel's PIN tool~\cite{pin} with models for the core, memory hierarchy, and on-chip networks. Area and power estimates are obtained from CACTI 6.5~\cite{cacti} using its most advanced technology node, 32nm. We use the SPEC CPU2006~\cite{spec} workloads with reference inputs. Furthermore, we use multiple inputs per workload to evaluate performance, energy, and area. We use SimPoint~\cite{simPoints} to choose a single representative region of 1 billion instructions in each application.

The key microarchitectural parameters are presented in Table~\ref{microParam}, with all core designs being two-wide superscalar with 64-entry instruction window and a cache hierarchy employing hardware prefetchers. We compare the MLP, performance, energy, and area overheads for the following four core designs:

\vspace{.02in}

\runinsec{In-order Core} We use an in-order stall-on-use core, resembling ARM Cortex-A7~\cite{cortexA7}, as a baseline.

\runinsec{Load Slice Core} LSC, as proposed by Carlson et al.~\cite{LSC}, with strict in-order memory slice execution. We model the A-IQ and B-IQ as each having 64 entries.

\runinsec{Freeway} Our proposed design with dependent slice tracking and a FIFO yielding queue (Y-IQ) to enable out-of-order execution among slices. To keep the total number of instruction queue entries same as in LSC, we model A-IQ, B-IQ and Y-IQ with 64-entries, 32-entries, and 32-entries respectively.

\runinsec{Ideal-sOoO} LSC with a fully out-of-order B-IQ. This design provides an upper bound on the performance limits of MLP in sOoO cores as it can execute MLP generating instructions from anywhere in the B-IQ, thus preventing stalled slices from blocking MLP exploitation. 

\runinsec{Out-of-Order Core} We use a fully out-of-order core, resembling ARM Cortex-A9, as MLP+ILP limit on performance.

\begin{figure*}[t]
    \centering
    \includegraphics[width=\textwidth, trim=15 150 15 140, clip]{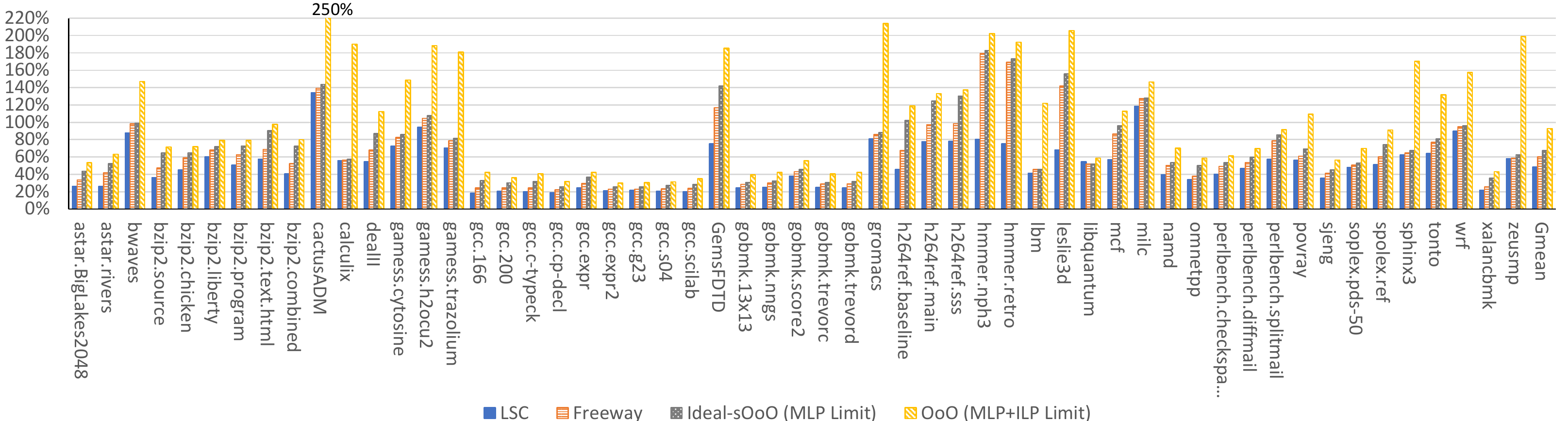}
    \vspace{-0.2in}
    \caption{Performance gain of different core designs over an in-order core.}\label{fig:performance}
    \vspace{-0.1in}
\end{figure*}

\section{Evaluation}
\label{sec:eval}

\subsection{Performance}
\label{sec:perf}
Figure~\ref{fig:performance} presents the performance gains of LSC and Freeway over the baseline in-order core. The figure also shows the performance limits of Ideal-sOoO (MLP limit) and full OoO (MLP+ILP limit) execution. The geometric mean (Gmean) performance difference between Freeway and LSC is 12\% as Freeway attains 60\% speedup over the in-order execution compared to 48\% for LSC.  
More importantly, Freeway is within 7\% of the performance of Ideal-sOoO, which is the upper bound on the performance achievable via MLP exploitation (MLP limit). An OoO core delivers 33\% more performance gain than Freeway, over an InO core, as it exploits both ILP and MLP, whereas Freeway targets only MLP. However, this additional performance comes with high area and power overheads (Section~\ref{sec:overheads}).

\runinsec{Freeway vs LSC} On individual workloads, Freeway performs significantly better than LSC on workloads such as \texttt{hmmer}, \texttt{leslie3d}, and \texttt{GemsFDTD} where dependent slices stall the B-IQ of LSC for a significant fraction of execution time (Figure~\ref{fig:stallDist}). Freeway eliminates these stalls by steering the dependent slices to the newly added Y-IQ, thereby executing the subsequent independent slice without stalling and boosting performance.

A closer inspection reveals that Freeway's performance advantage over LSC is a function of not only the number of stalls caused by dependent slices, but also where producer slices hit in the memory hierarchy. For example, workloads such as \texttt{gcc}, \texttt{soplex}, and \texttt{omnetpp}, where slice dependencies stall execution for more than 45\% of time in LSC, benefit only moderately from Freeway. The reason for this behaviour is that more than 70\% of the producer slices in these workloads miss in the on-chip cache hierarchy and must be loaded from memory. (Figure~\ref{fig:sliceDepStalls}). This latency stalls instruction retirement, and therefore causes the instruction window to fill, which blocks further MLP generation and limits performance. In contrast, Freeway delivers significantly more performance (95\% and 76\%, respectively) for \texttt{hmmer} and \texttt{leslie3d}, despite LSC stalling on slice dependencies for only 30\% of the execution time. This is because almost all of the producer slices in these workloads hit in the L1 cache. Therefore, the instruction window is rarely full and Freeway can continuously exploit MLP and improve performance.

Finally, Figure~\ref{fig:performance} also shows that both Freeway and LSC perform similarly on workloads such as \texttt{zeusmp}, \texttt{milc}, \texttt{lbm}, and \texttt{calculix}. These workloads do not have many dependent slices and the corresponding stalls, as shown in Figure~\ref{fig:stallDist}, are minimal. As a result, Freeway does not have much opportunity for improvement over LSC.

\runinsec{Freeway vs Ideal-sOoO vs OoO} Figure~\ref{fig:performance} shows that the majority of the benefits of full OoO execution can be obtained primarily by exploiting MLP as Ideal-sOoO (MLP limit) achieves about 72\% of the performance benefits of full OoO (MLP+ILP) execution. Furthermore, the figure also shows that Freeway captures the bulk of this MLP opportunity and reaches within 7\% of the performance delivered by idealized MLP extraction (Ideal-sOoO design). 
Compared to OoO execution, which targets both ILP and MLP, Freeway falls short on workloads that present significant ILP opportunity, such as \texttt{calculix}, \texttt{gromacs}, \texttt{zeusmp} etc., as it exclusively aims for MLP. However, on workloads that offer little ILP, such as \texttt{sjeng}, \texttt{perlbench}, \texttt{xalancbmk}, etc., Freeway is within 15\% of the full OoO performance. Overall, full OoO execution provides 93\% performance gain over in-order execution compared to the 60\% gain of Freeway. However, the additional performance of OoO comes at a significantly higher area and power costs as discussed in Section~\ref{sec:overheads}.

\runinsec{Performance impact of load speculation} For the results presented in Figure~\ref{fig:performance}, load instructions are issued only when all the older store instructions have their addresses available and there is no aliasing. Though such conservative scheduling reduces the complexity of load-store unit, it also loses performance opportunities. To understand the performance opportunity in load speculation for different core designs, we assume an oracle memory dependence predictor and issue a load if it is not predicted to be aliased with older stores even if their addresses are not yet computed. Notice that, in addition to OoO core, Freeway also benefits from load speculation. This is because, as discussed in Section~\ref{sec:mem_order}, a load in Freeway might have its address available, i.e., it’s ready for execution, before all the older stores compute their addresses. Finally, Ideal-sOoO core would also benefit from load speculation as its B-IQ is fully out-of-order. The in-order core and LSC, in contrast, do not benefit from load speculation because they compute all load and store addresses in program order. Our results shows that load speculation enables Freeway, Ideal-sOoO, and OoO cores to provide 63\%, 70\%, and 97\% performance gain over in-order core; whereas their respective performance gains without load speculation are 60\%, 67\% and 93\%. On individual applications, the performance gain from load speculation is as high as 20\%, for example on bzip2.program. These results suggest that load speculation helps Freeway, Ideal-sOoO, and OoO about equally. For the rest of the evaluation, we use the conservative scheduling policy, i.e., without load speculation.

\subsection{Analysis of the Remaining Opportunity}
\label{sec:anaRemOpp}

Figure~\ref{fig:performance} shows that despite its dependence aware slice execution, Freeway lags behind the optimal performance achievable via MLP exploitation (``Ideal-sOoO (MLP limit)") by 7\%. We observe that there are two main factors that cause this gap: First, Freeway addresses only the slice dependence related stalls but not the other stall sources such as Load-Store Aliasing, Empty B-IQ, etc. (described in Section~\ref{sec::stallDist}). Second, buffering all dependent slices in a single Y-IQ leads to stalls when a younger slice becomes ready earlier than an older slice, although this is infrequent. 
Here we analyze the performance loss due to these factors and explore potential solutions.

As Figure~\ref{fig:stallDist} shows, \textit{Empty B-IQ} is the largest source of stalls after slice dependence stalls. However, mitigating them requires improvements in core components other than instruction scheduling. For example, it either requires a better front-end, if the B-IQ is empty due to branch mispredictions or instruction cache misses, or it requires a larger instruction window, if the window is full when the B-IQ is empty. As the focus of this work is instruction scheduling, we do not explore techniques to mitigate these stalls.

\begin{figure*}[t]
    \centering
    \includegraphics[width=0.95\textwidth, trim=95 200 60 180, clip]{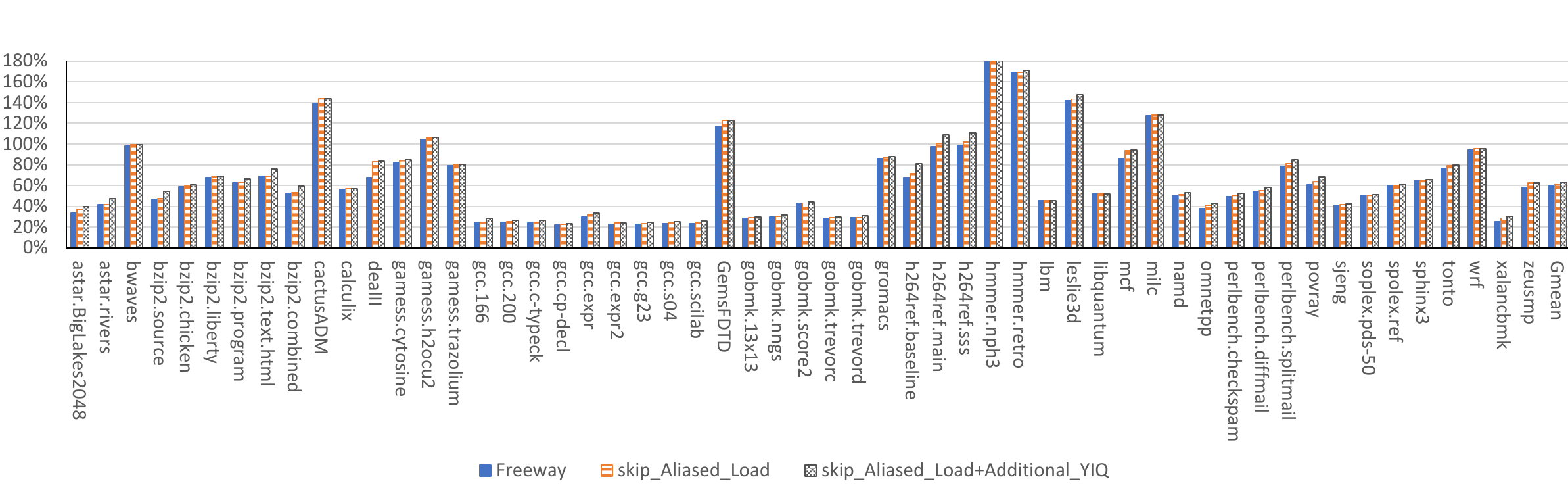}
    \vspace{-0.15in}
    \caption{Freeway performance gain achieved by skipping the aliased loads and adding another Y-IQ.}\label{fig:anaRemOpp}
    \vspace{-0.2in}
\end{figure*}

The next largest source of stalls, as Figure~\ref{fig:stallDist} shows, is the stalls coming from load-store aliasing. Such load-store aliasing causes LSC to stall for 8\% of the execution time. Furthermore, as Freeway enables early execution of independent slices, it can potentially expose more aliasing if some of the aliased loads were earlier hidden behind the dependent slices in the B-IQ of LSC. To quantify the resulting performance loss, we simulate skipping the aliased loads and issuing the subsequent instructions if they are ready. Though impractical, such scheduling shows the potential benefits of eliminating the load-store aliasing related stalls. The \textit{skip\_Aliased\_Load} bar in Figure~\ref{fig:anaRemOpp} shows that Freeway obtains only 2\% additional performance by eliminating all such stalls. As the \textit{Other} stall sources contribute even less, we do not quantify their impact on performance loss. From this analysis we see that even completely addressing the load-store aliasing related stalls would result in little performance gain.

As discussed in Section~\ref{sec::whereToBuffer}, buffering all dependent slices in a single Y-IQ might lead to stalls if younger slices become ready before the older ones (either the younger slices are at a lower dependence depth or their producers hit closer to the core in the memory hierarchy).  As almost all producers hit in the L1 cache (Figure~\ref{fig:prodSliceHitSite}), we only consider mitigating stalls due to slice dependence depth by adding a single additional Y-IQ. Here we explore the benefits of having the first Y-IQ buffer only the dependent slices at dependence depth 1, while the rest of the dependent slices go to a second Y-IQ. As a result, the stalls in the first Y-IQ will be reduced as all the slices are at the same dependence depth. The \textit{skip\_Aliased\_Load+Additional\_Y-IQ} bar in Figure~\ref{fig:anaRemOpp} shows that adding a second Y-IQ brings only 1.5\% additional performance. These results also confirm our hypothesis that a single Y-IQ is enough to capture the most of the opportunity in out-of-order slice execution.

If combined, the above optimizations would bring performance to within 3.5\% of the optimal. However, individually they provide only minimum performance returns on the resource investment. 

\vspace{-0.1in}

\subsection{Breaking Down the Performance Contribution of Different MLP Generators}
\label{sec::MLPBreakdown}

Our proposed core design, Freeway, generates MLP by enabling slices to bypass the stalled non-slice instructions and enabling independent slices to bypass the stalled dependent ones. In addition, we employ a LLC prefetcher that increases the number of overlapping memory requests, hence MLP, by speculatively generating accesses for cache blocks that are likely to be required soon. This section quantifies the contribution of each of these components towards overall performance.

\begin{figure*}[t]
    \centering
    \includegraphics[width=\textwidth, trim=45 150 50 125, clip]{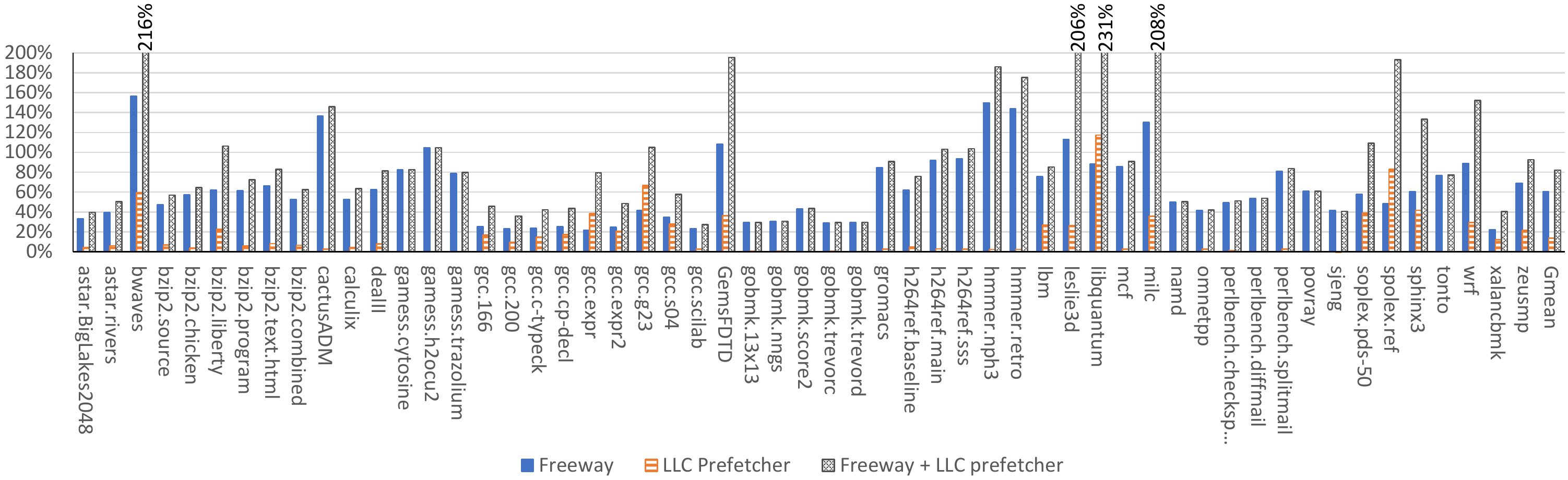}
    \vspace{-0.25in}
    \caption{Performance breakdown: gains achieved individually by Freeway and LLC prefetcher, and their combined gain over a prefetcherless in-order core. (The baseline core in all other studies does include a LLC prefetcher.)}\label{fig:pFetchStudy}
    \vspace{-0.25in}
\end{figure*}

Figure~\ref{fig:pFetchStudy} presents the performance gain achieved by the individual MLP generating components, i.e. Freeway and the LLC prefetcher, as well as the gain when they work together. For this study, we do not model prefetcher in the baseline in-order core to help evaluate the performance gains of the LLC prefetcher. In all other studies, however, our baseline in-order core (and all other evaluated cores including Freeway) does include a LLC prefetcher. 

As the figure shows, Freeway provides a much higher performance gain (61\%) than the LLC prefetcher (14\%). This is because Freeway hides L1, LLC, and to some extent memory latency by generating parallel memory accesses and overlapping their latency. LLC prefetcher, in contrast, hides only the latency between LLC and memory, however the LLC (and even L1) latency is still exposed due to the serial accesses up until LLC. We also evaluated aiding the LLC prefetcher with an L1 prefetcher (stride prefetcher with 4 independent streams); however, it provides only about 1.5\% additional Gmean performance.

\begin{figure*}[t]
    \centering
    \includegraphics[width=0.95\textwidth, trim=55 320 50 360, clip]{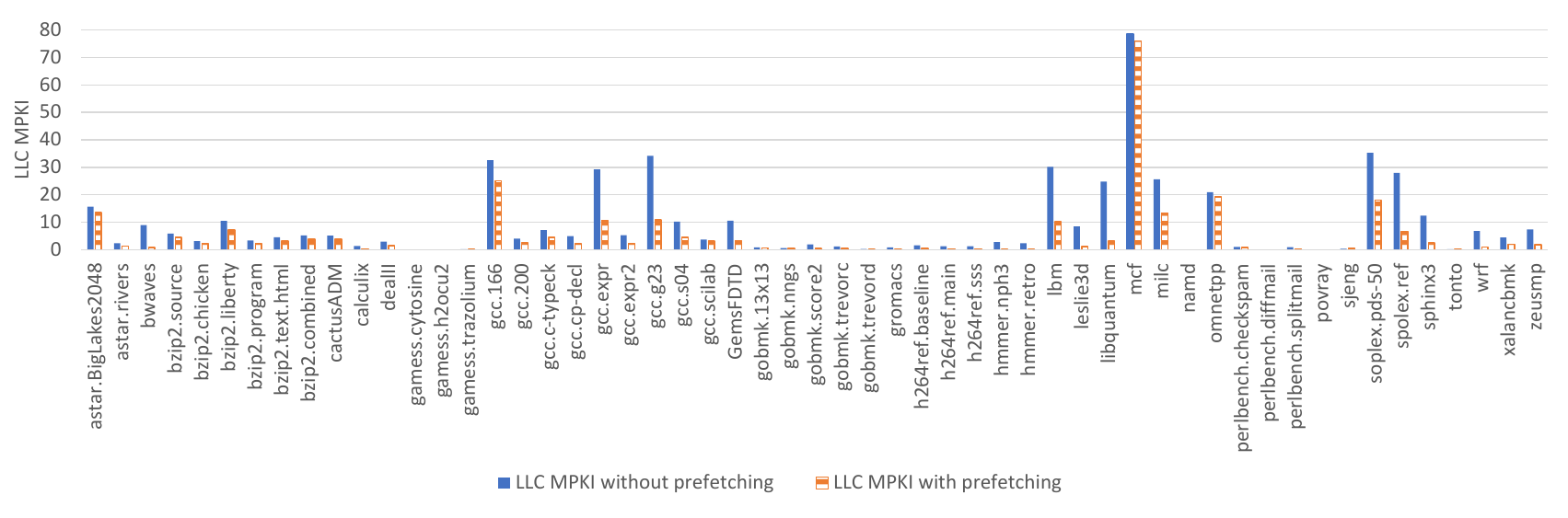}
    \vspace{-0.15in}
    \caption{LLC MPKI (misses per kilo instructions) without and with LLC prefetcher.}\label{fig:missRates}
    \vspace{-0.2in}
\end{figure*}

Looking closely at individual workloads, there are three applications, \texttt{gcc} (expr and g23 inputs), \texttt{libquantum}, and \texttt{soplex} (ref input), where the LLC prefetcher outperforms Freeway. Figure~\ref{fig:missRates} shows that all of these applications have high LLC MPKI (25-34) and thus get most of their data from memory. The LLC prefetcher delivers higher performance due to their prefetch friendly access pattern as it is able to reduce the LLC MPKI to 3-11. In contrast, Freeway is not able to generate enough MLP mainly due to a combination of two factors. First, the high LLC MPKI leads to frequent full instruction window stalls (long memory latency on LLC misses causes instruction window to fill up) which block further MLP generation. Second, and more importantly, as shown in Figure~\ref{fig:sliceDepth}, these applications have a larger number of dependent slices which cannot be executed until their parent slices receive data, thus further lowering MLP.

For the other benchmarks with high LLC MPKI, Freeway outperforms the LLC prefetcher either because the benchmarks do not have many dependent slices, and therefore Freeway generates enough MLP, or their access patterns are not prefetch friendly and, therefore the LLC prefetcher does not perform well. For example, \texttt{lbm} does not have any dependent slices and, therefore, Freeway generates significant MLP. Conversely, \texttt{mcf}'s access pattern is not prefetch friendly, therefore, resulting in poor LLC prefetcher performance.

Another important finding from Figure~\ref{fig:pFetchStudy} is that combining Freeway with LLC prefetcher provides considerably higher performance gain (82\%) than that delivered individually by Freeway (61\%) and the LLC prefetcher (14\%). This is because, when combined, more memory accesses generated by Freeway are served from LLC due to the prefetched blocks, thereby reducing average memory access time. In addition, if these accesses were causing full instruction window stalls by blocking the instruction commit, their early completion will also enable subsequent memory accesses to enter the instruction window and execute early, thus further improving MLP and performance.

Overall the results in Figure~\ref{fig:pFetchStudy} show that MLP generated by Freeway provides significantly higher performance than the LLC prefetcher. These results also imply that if designers must pick one technique among several competing candidates for MLP generation, for example in area-constrained designs, it is better to invest real estate in Freeway rather than in LLC prefetcher.

\subsection{Freeway's Efficacy at Smaller Instruction Queues}
\label{sec::Qsizesens}

\begin{figure*}[t]
    \centering
    \includegraphics[width=\textwidth, trim=30 150 30 125, clip]{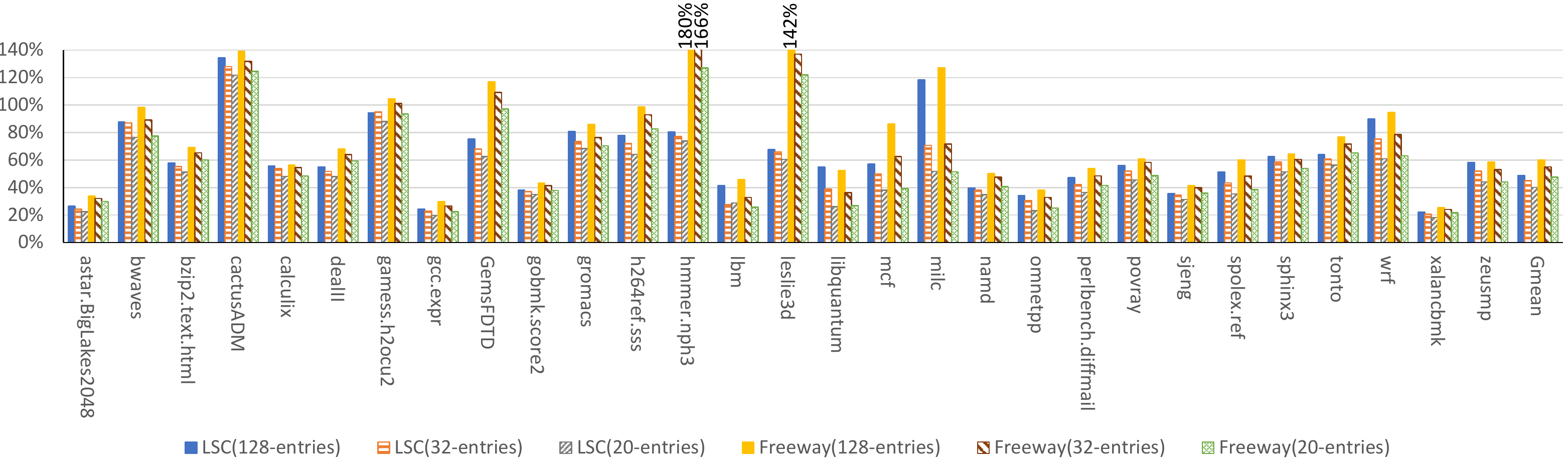}
    \caption{Performance gain with different number of instruction queue entries. The ratio of entries among queues is 1:1 (A-IQ:B-IQ) in LSC and 2:1:1 (A-IQ:B-IQ:Y-IQ) in Freeway. (For readability, we show only one input per benchmark)}\label{fig:QSizeSen}
    \vspace{-0.2in}
\end{figure*}

As Section~\ref{sec:perf} detailed, Freeway performs significantly better than LSC when both are given equal number of instruction queue entries (128).  
This section investigates the designs' relative sensitivity to reducing the number of instruction queue entries and explores how few entries Freeway can work with while still delivering similar performance to LSC.

Figure~\ref{fig:QSizeSen} presents the performance gains achieved by LSC and Freeway with a total of 128, 32 and 20 instruction queue entries. We observed that the performance with 128 and 64 entries is similar (\textless 1\% difference), therefore we do not show 64-entry design point in the figure. The ratio of entries among different queues remains same as in the 128-entry case i.e. 1:1 (A-IQ:B-IQ) in LSC and 2:1:1 (A-IQ:B-IQ:Y-IQ) in Freeway. The results show that Freeway always performs better than LSC when they both feature equal number of queue entries regardless of whether it's 128, 32, or 20-entries. Furthermore, Freeway with just 32 entries provides about 6\% more gain than 128-entry LSC, over InO core, and Freeway requires barely 20 instruction queue entries to provide similar performance to that of the 128-entry LSC. 

Looking at individual benchmarks, there is a set of applications where the 20-entry Freeway performs significantly better the 128-entry LSC. For example, Freeway performs about 50\% better on \texttt{hmmer} and \texttt{leslie3d} and about 20\% better on \texttt{GemsFDTD}. These are the applications where letting younger independent slices to bypass the older dependent ones offers significant performance opportunities as shown in Figure~\ref{fig:opportunity}. The results in Figure~\ref{fig:QSizeSen} imply that even a small Y-IQ, with only 5 entries, enables significant number of independent slices to bypass the dependent ones, thus providing a high performance gain. Whereas a larger B-IQ in LSC does not provide additional performance as it simply results in more independent slices sitting behind the stalled dependent ones.

However, the 20-entry Freeway performs worse than the 128-entry LSC for some applications like \texttt{milc}, \texttt{wrf}, \texttt{zeusmp} etc. These applications do not present much performance opportunity as shown in Figure~\ref{fig:opportunity}. In the absence of performance opportunity, the frequent dispatch (insertion into instruction queues) stalls caused by smaller instruction queues result in 20-entry Freeway performance dropping below that of the 128-entry LSC. The dispatch stalls are more frequent with smaller queues as they fill up quickly and the dispatch stage stalls as soon as any one of the queues is full and the next instruction also needs to be dispatched to that same queue.  
Though the combination of low opportunity and frequent dispatch stalls causes a performance drop in these applications, Freeway still performs slightly better than LSC when both are given same number of queue entries.

Overall, the results in Figure~\ref{fig:QSizeSen} show that Freeway can tolerate more than 6x reduction in the instruction queue entries (128 to 20) while delivering similar performance as 128-entry LSC due to its dependence aware slice execution.

\subsection{Sensitivity to L1 Cache Latency}
\label{sec::L1LatSens}

\begin{wrapfigure}{R}{0.65\textwidth}
    \centering
    \includegraphics[width=0.65\textwidth, trim=60 90 90 90, clip]{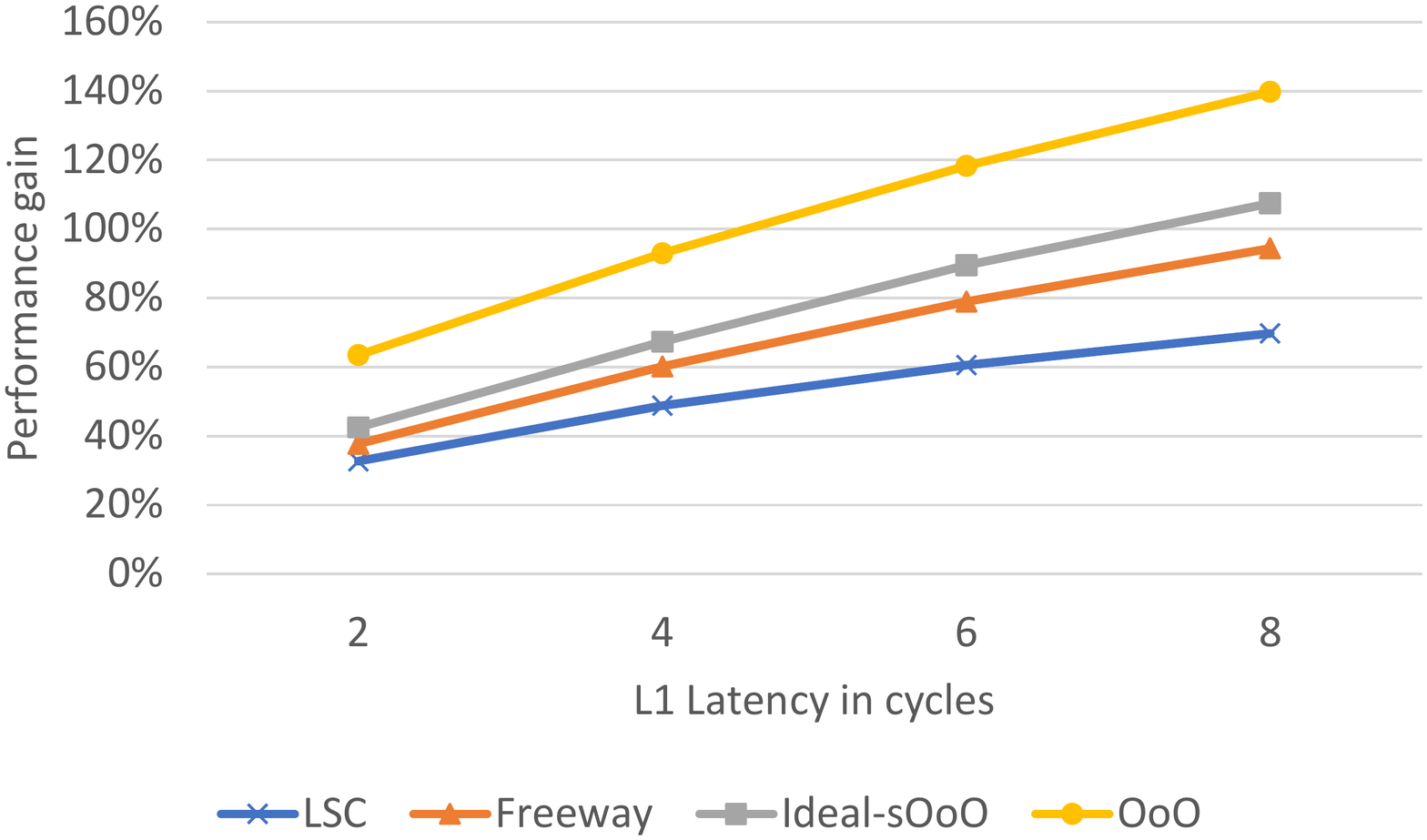}
    \vspace{-0.2in}
    \caption{Performance gain at different L1 cache latencies.}\label{fig:l1dLatSens}
    \vspace{-0.1in}
\end{wrapfigure}

As the majority of stalls in LSC occurs while producer slices fetch data from L1 cache (Figure~\ref{fig:sliceDepStalls}), we study the sensitivity of Freeway's performance gain to the L1 cache latency.
In addition, we also study how close Freeway comes to the Ideal-sOoO and OoO cores at different L1 latencies. For this study, we vary the L1 latency (load-to-use) from 2 to 8 cycles. The default L1 latency used for the other experiments is 4 cycles.

Figure~\ref{fig:l1dLatSens} presents the performance gain of different cores over an in-order core as a function of L1 latency. The first point to notice is that the gain of all cores is smaller at lower latencies. For example, Freeway delivers 60\% performance gain over in-order core at 4-cycle latency, whereas it drops down to 38\% at 2-cycle latency. This is because the baseline in-order core itself performs increasingly better at lower latencies as the dependent instructions stall for less time on L1 hits. As a result, the performance opportunity, and the achieved gain, for other cores is reduced.

\begin{figure*}[t]
    \centering
    \includegraphics[width=\textwidth, trim=90 200 60 190, clip]{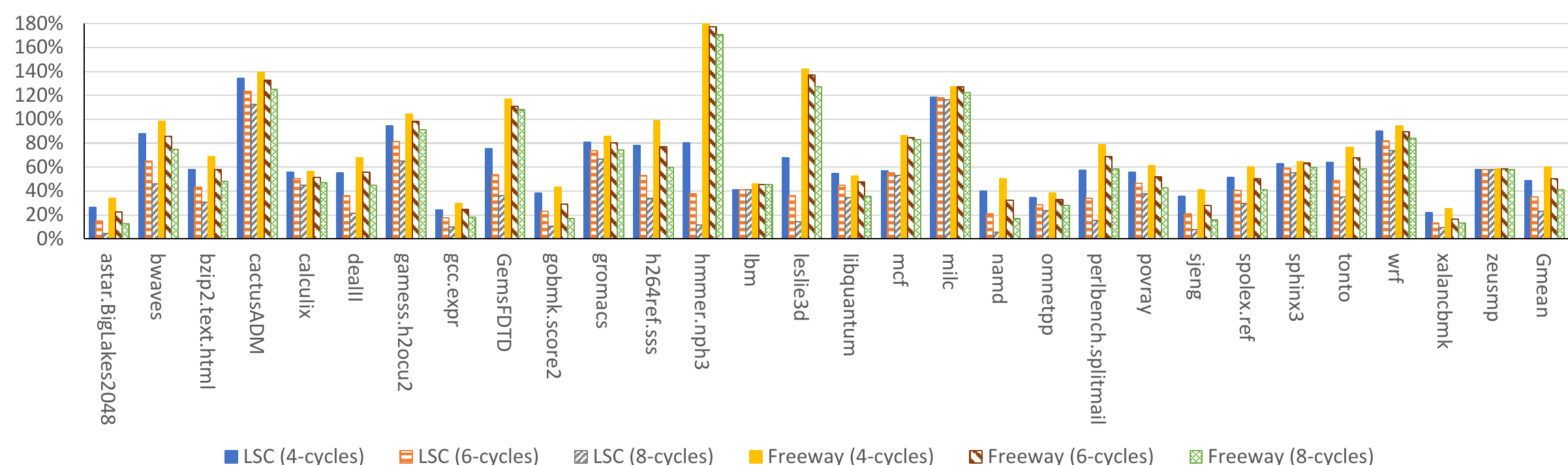}
    \vspace{-0.1in}
    \caption{Freeway and LSC performance gain at different L1 cache latencies while the baseline in-order core L1 latency fixed at 4 cycles. (For readability, we show only one input per benchmark)}\label{fig:l1dLatSens_FreewayLSC}
    \vspace{-0.1in}
\end{figure*}

An important finding from Figure~\ref{fig:l1dLatSens} is that even though the performance gap between Freeway and LSC is very small (5\%) at 2-cycle latency, it widens quickly at larger latencies: 12\% at 4-cycle, 18\% at 6-cycle, and 25\% at 8-cycle latency. This is because the stalled dependent slices prevent the subsequent independent slices from executing for longer intervals at larger latencies in LSC, thereby limiting MLP. In contrast, Freeway keeps executing independent slices from B-IQ while the dependent ones wait longer in the Y-IQ. Overall, these results demonstrate that Freeway is better than LSC in tolerating higher L1 latencies.

The figure also shows that the performance gap between Freeway and Ideal-sOoO (which represents the MLP limit) grows at a significantly lower rate compared to the gap between Freeway and LSC. This gap increases from about 5\% at 2-cycle L1 latency to 13\% at the 8-cycle latency. As Ideal-sOoO executes slice instructions fully out-of-order, it is able to cover the MLP opportunities missed by Freeway, as discussed in Section~\ref{sec:anaRemOpp}. The performance gap between Ideal-sOoO and fully OoO also increases while going from 2-cycle to 8-cycle latency because OoO can hide the increased latency better by executing more ILP-generating instructions.

\noindent
\textit{\underline{Freeway vs LSC  (with fixed L1 latency for the baseline core):}}
To understand how much additional L1 latency Freeway can tolerate compared to LSC, we keep the baseline in-order core's L1 latency constant at 4 cycles (the default value) and vary LSC and Freeway L1 latency to 4, 6, and 8 cycles. Notice that this is in contrast to the results in Figure~\ref{fig:l1dLatSens} where we vary the L1 latency for the baseline in-order core along with other core designs. 

The results in Figure~\ref{fig:l1dLatSens_FreewayLSC} show that Freeway not only performs better than LSC while considering equal L1 latency for both but also when it has to tolerate higher latency than LSC. For example, LSC delivers about 48\% performance gain over the baseline InO core with a 4-cycle L1 latency, whereas Freeway achieves 50\% gain even at a higher latency of 6 cycles. Similarly, LSC provides 35\% gain at 6-cycle L1 latency while Freeway attains 41\% gain even with 8-cycle latency. These results further validates Freeway's advantage over LSC in tolerating L1 latency.

These results also imply that Freeway is likely to benefit more from a larger L1 cache (with correspondingly higher access latency) than LSC because of its better L1 latency tolerance.

\vspace{-0.1in}
\subsection{Scalability Analysis}
\label{sec::WindowSizeSens}

The instruction window size, e.g. scoreboard size in Freeway, limits the number of in-flight instructions. Consequently, the number of outstanding memory accesses, hence MLP, is also bounded by the size of instruction window. While prior work~\cite{LSC, freeway} evaluated sOoO cores only at small instruction windows, we analyze their scalability across the full spectrum of instruction window widths and depths (i.e. from wimpy to brawny cores) by varying the instruction window size from 32 to 224 entries. We also size other structures (instruction queues, execution units, etc.) appropriately in accordance with the instruction window size. Furthermore, we use an issue width of 2 for the instruction windows of 32- and 64-entries, while 128-, 168-, 192-, and 224-entry instruction windows use issue widths of 3, 4, 6, and 8 respectively.

\begin{figure}[t]
    \centering
    \includegraphics[width=0.7\textwidth, trim=60 270 50 310, clip]{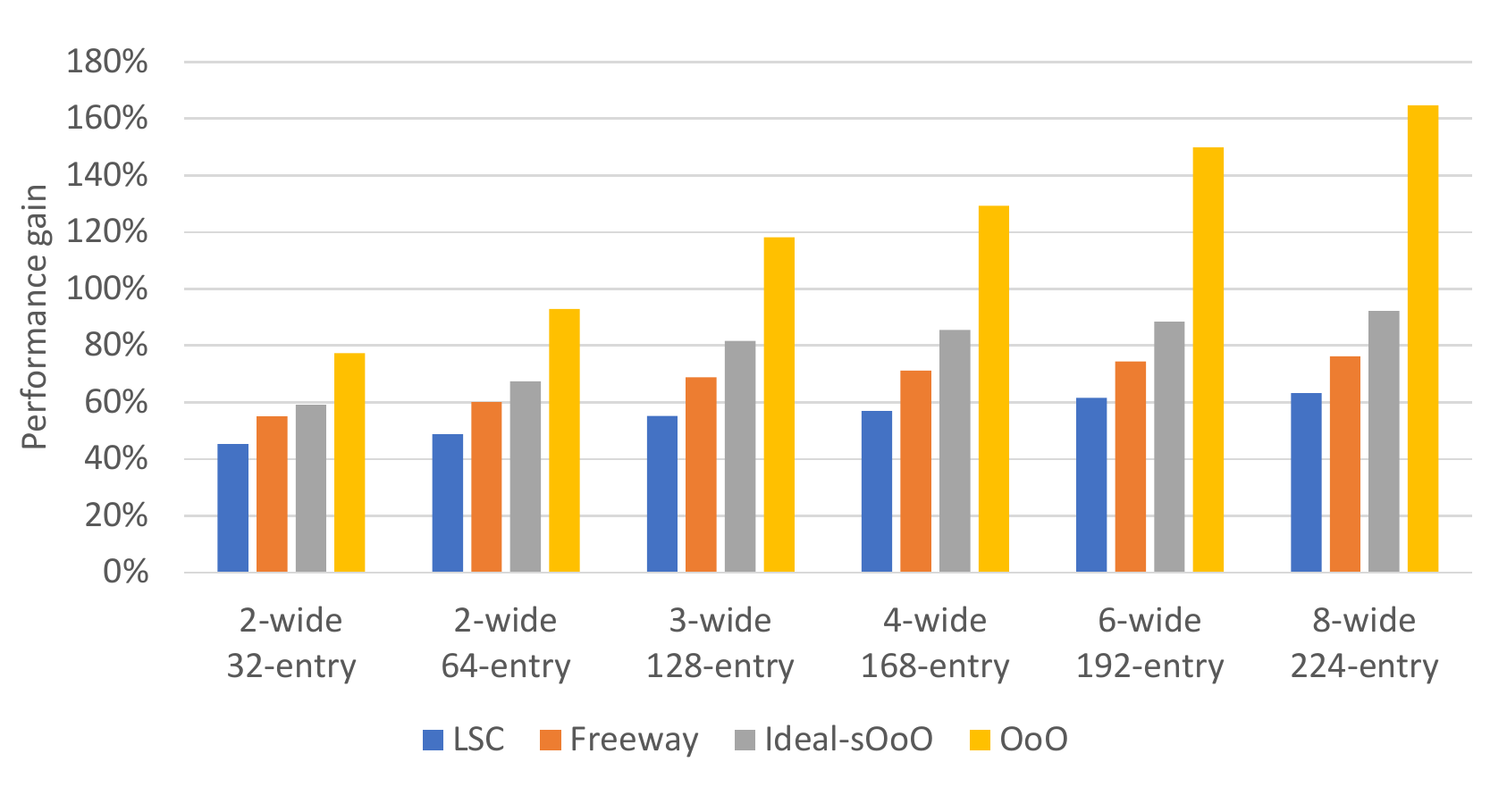}
    \vspace{-0.2in}
    \caption{Performance sensitivity to core size. x-axis are labeled with issue-width (x-wide) and instruction window size (y-entry).}\label{fig:sensitivityInstWin}
    \vspace{-0.2in}
\end{figure}

The performance results (geometric mean) across the benchmark suite are presented in Figure~\ref{fig:sensitivityInstWin}. The results show that Freeway's performance advantage over LSC increases from about 10\% to 14\% while moving from 32 to 224 entry instruction window. Their performance difference at the 32-entry window is small because there are not many MLP opportunities in such a small window due to fewer in-flight instructions. However, as the number of in-flight instructions increases with window size, Freeway exposes increasingly more MLP by executing independent slices unobstructed, whereas in LSC most of the additional slices simply sit behind the stalled ones. The performance gap between Freeway and Ideal-sOoO also increases with window size, though Freeway is still within 16\% of Ideal-sOoO (limits of MLP) even at a large 8-wide 224-entry instruction window. 

The performance gains of fully out-of-order execution increase even more with the larger windows due to its ability to utilize both MLP and ILP. An important finding from the results of Figure~\ref{fig:sensitivityInstWin} is that MLP contributes less to overall performance at large instruction windows. For example, Ideal-sOoO core, which exploits the maximum MLP, is within 18\% of full OoO performance at a small 32-entry instruction window. However, the difference grows with instruction window size. This is because larger instruction windows offer more ILP opportunities to OoO execution, thus widening the performance gap with Ideal-sOoO which only extracts MLP.

These results highlight that further research is required for scaling sOoO cores to larger instruction windows. Furthermore, given that MLP's contribution to overall performance reduces at large windows and that Freeway already extracts a large portion of the available MLP, the research in scaling sOoO cores should focus on ILP rather than capturing the MLP opportunities missed by Freeway.

\vspace{-0.1in}

\subsection{Area and Power overheads}
\label{sec:overheads}

To evaluate the area and power overheads of different core designs, we use CACTI 6.5 to compute the area and power consumption of each of their major components in 32nm technology. The area overhead of LSC is about 15\% over the baseline in-order core. Freeway requires very little additional hardware over LSC: one bit per entry in RDT, 7 bits per entry in the store buffer, and the Y-IQ and logic to issue instructions from it. As a result, it needs only 1.5\% additional area. In contrast, as reported by~\cite{LSC}, the OoO core incurs an area overhead of 154\% over the baseline in-order core.

For power calculations, we use static power consumption and per-access energy values from CACTI and combine them with activity factors obtained from the timing simulations to compute power requirements of each component. Our evaluation, together with prior results~\cite{LSC}, show that LSC, Freeway, and OoO core increase the average power consumption by 1.22x, 1.24x, and 12.6x, respectively, over an in-order core. These results demonstrate the exorbitant area and power costs of the moderate performance benefits achieved by OoO core over more efficient slice-out-of-order core designs such as Freeway.

\section{Related Work}
\label{sec:related}

\subsection{Tolerating Long Memory Latency}
Existing research on tolerating memory access latency to prevent cores from stalling can be divided into two broad categories: techniques that \textit{extract MLP} by overlapping the execution of multiple memory requests, and techniques that \textit{prefetch data} proactively into caches by predicting future memory addresses. To some extent, these categories are complementary as prefetching can increase the number of overlapping memory requests by speculatively generating accesses that would otherwise be serialized due to dependencies or lack of resources. However, the energy efficiency of the majority of these techniques is bounded by the underlying energy intensive OoO core. Freeway, in contrast, provides an energy efficient alternative that these techniques can build on to potentially raise overall efficiency.

\vspace{.02in}

\runinsec{MLP Extraction} OoO execution is the most generic approach for generating MLP. However, it is limited by the instruction window size. The following techniques break this size barrier to boost MLP extraction:

\noindent
\textit{\underline{Runahead Execution:}} Runahead Execution~\cite{runahead} improves MLP by pre-executing instructions beyond a full instruction window. Once the OoO core stalls due to a full ROB, Runahead checkpoints the processor state, tosses out the ROB stalling instruction, and continues to fetch subsequent instructions. These new instructions are executed if their source data is available, thereby generating additional memory accesses and boosting MLP. Recent work has improved several aspects of runahead execution. Filtered runahead~\cite{filteredRunahead} clock- and/or power-gates the core front-end while supplying instructions from a buffer to reduce energy consumption. Precise Runahead~\cite{preciseRunahead} reduces the overhead of restarting regular execution after coming out of the runahead mode. Continuous Runahead~\cite{continuousRunahead} continuously executes MLP generating instructions in parallel with core, instead of waiting for the core to stall, to improve runhaed coverage. 

\noindent
\textit{\underline{Helper Threads:}} These techniques rely on pre-executing ``helper threads" or code segments to generate MLP. A helper thread is a stripped down version of the main thread that only includes the necessary instructions to generate memory accesses, including control flow instructions. However, helper threads require an independent execution context (SMT or a CMP core) for their execution. As they generate memory accesses in parallel with the main thread, they increase MLP and/or prefetch data.

Helper threads can be generated either in software or dynamically in hardware. On the software side, many prior works have proposed compiler/programmer driven approaches for helper thread generation~\cite{specComp, compAlgo, softControl, execBased, interCore, studyComp, DAE} while others have proposed dynamic compilation techniques~\cite{accelPreComp, sunHelper}. These techniques either execute the helpers threads on an available SMT context~\cite{specComp, studyComp} or require a dedicated core~\cite{interCore}. 

Collins et al.~\cite{dynSpecComp} explored helper thread generation in hardware by tracking dependent instruction chains in the back-end. To keep the helper thread generation off the critical path, they introduced large, post-retirement, hardware structures to filter the desired instructions. Once the helper threads were generated, they were stored in a large cache and run on a free SMT context. Annavaram et al.~\cite{depGraph} also extracted the dependent chains of operations that were likely to result in a cache miss in hardware, though from the front-end during instruction decode, and added a dedicated back-end for the execution of such chains.

To summarize, helper threads incur significant overhead as they require: 1) an independent execution context (SMT, a CMP core, or dedicate hardware) for their execution, 2) a mechanism to construct them either in hardware or software, 3) duplicated instruction execution in the main thread and helper thread.

\vspace{.02in}

\runinsec{Prefetching} 
Prefetchers predict future addresses based on prior memory access patterns. However, they either have limited coverage due to being limited to simple access patterns or require extensive hardware. For example, stride and stream prefetchers~\cite{prefetchBuff, streamBuff} require only simple hardware but are limited to regular access patterns. Advanced prefetchers, such as Correlation Prefetchers~\cite{markov, deadBlock, linkedDC}, enable complex access pattern prefetching at the cost of large tables to link the past miss addresses to future miss addresses. Spatial and temporal streaming based prefetching has also been explored in server domain~\cite{sms, tms, stems}, though they still incur significant storage and energy overhead. Recently, co-design of prefetching with replacement policies has also been explored in~\cite{PACMan, killPC}.

\subsection{Energy-efficient Core Design} 

Instruction scheduling is one of the most energy hungry operations in modern OoO cores~\cite{complexityEffective}. Therefore, researchers have proposed a number of techniques to reduce its energy requirements. Recent research in this domains exploits two properties, instruction readiness and instruction criticality, to minimize scheduling energy overhead. 

Shioya et al.~\cite{fxa} observed that a large fraction to total dynamic instructions is either ready for execution at dispatch stage or becomes ready within a few cycles of dispatching to the issue queue. These instructions do not benefit from OoO scheduling as they would execute without stalls even in an in-order pipeline. Therefore, they propose an architecture that attempts to execute all instructions via in-order pipeline stages before dispatching the unexecuted ones to OoO pipeline, thereby reducing scheduling energy. FIFOrder~\cite{fiforder}, instead of trying to execute all instruction via in-order stages, dispatches ready instructions to a FIFO issue queue and non-ready instructions to an OoO (content addressable memory based) issue queue. As the OoO queue handles fewer instructions, FIFOrder reduces its depth and width, thus reducing the scheduling energy cost. Another recent architecture, CASINO core~\cite{casino}, also targets ready instructions to simplify instruction scheduling.

Other researchers have targeted instruction criticality to cut the energy cost of scheduling. For example, Long Term Parking (LTP)~\cite{LTP}, at dispatch stage, allocate issue queue entries only to critical instructions, whereas non-critical instructions are \textit{parked} in a FIFO parking queue. Parked instructions are allocated issue queue entries only when they reach close to the head of reorder-buffer. As the parking queue reduces pressure on issue queue, LTP reduces its dimensions to reduce its energy requirements. A recent work, Delay and Bypass~\cite{dnb}, advocates exploiting both readiness and critically simultaneously to further improve energy savings.

Though all of these designs reduce instruction scheduling energy, they still feature CAM based instruction queues, albeit smaller than OoO cores. Therefore, their energy savings are not as high as those of sOoO cores. A recent design, Forward Slice Core(FSC)~\cite{fsc}, is the closest approach to sOoO cores in that it builds on a stall-on-use in-order core and uses only FIFO queues for instruction scheduling. Unlike sOoO cores, which specifically target MLP, Forward Slice Core (FSC) aims to extract generic ILP. Therefore, some of its features can be borrowed for ILP extraction in sOoO cores. Specifically, compared to sOoO cores, FSC enables non-slice instructions that do not depend on load instructions to bypass the load dependent instructions via a dedicated IQ, thereby extracting ILP among non-slice instructions. In sOoO cores, in contrast, as all non-slice instructions, whether load dependent or not, share the same IQ, they miss the ILP extraction opportunity. sOoO cores can borrow FSC’s mechanism to split load dependent and independent non-slice instructions into different queues to improve ILP extraction. However, this is only one of the many possible ways of splitting non-slice instructions. 
Other splitting criteria could include dispatching integer and floating-point instructions to different queues, creating and dispatching slices of branch instructions to dedicated queues to potentially reduce the branch misprediction penalty, or dispatching consumers of high vs. low latency loads to different queues, or splitting instructions based on fanout. Further investigation is needed to understand the trade-offs provided by these criteria.

FSC also shares some features with sOoO cores. For example, its mechanism to create forward slices is very similar to Freeway’s mechanism of identifying dependent slices. Also, as FSC does not explicitly identify address generating instructions (AGIs) and mixes them with non-AGIs in its Main Lane, load/store address generation is likely to be delayed, compared to sOoO cores, which would limit MLP extraction.

Besides instruction scheduling, register renaming is also an energy intensive operation~\cite{complexityEffective} which has received researchers' attention. Gonzalez et al.~\cite{virtreg} proposed to delay the register allocation until a late pipeline stage to reduce register file pressure, thus delivering same performance with a smaller and less energy hungry register file. Tabani et al.~\cite{tabani} proposed a technique that releases physical registers earlier than conventional techniques, thus reducing register file pressure, size, and energy requirements. A recent work, STRAIGHT~\cite{straight}, proposes an instruction set architecture that eliminates the register renaming altogether.

\section{Conclusion}
\label{sec:conc}

Tolerating long memory and LLC access latencies is critical for performance. MPL exploitation techniques such as out-of-order execution, runahead execution, etc., have been successful in hiding these latencies, however, at the cost of large energy overheads. Recent attempts to address these in an energy-efficient manner have led to \textit{slice-out-of-order} (sOoO) cores. These cores construct slices of MLP generating instructions and execute them out-of-order with respect to the rest of instructions. However, the slices and the remaining instructions, by themselves, still execute in-order. By limiting the out-of-order execution to only between slice and non-slice instructions, sOoO cores are able to achieve much of the MLP benefits of OoO processor with far less hardware overhead. 

This work introduces Freeway, a highly energy-efficient core that approaches the MLP benefits of full out-of-order execution. To keep the energy overhead low, Freeway builds upon a modern sOoO core. We show that, though energy-efficient, state-of-the-art sOoO cores miss significant MLP opportunities due to inter-slice dependencies. Freeway addresses this bottleneck by \textit{identifying} dependent slices and introducing an efficient \textit{dependence-aware} slice execution policy based on a detailed analysis of slice behaviour. Freeway's policy forces dependent slice to \textit{yield} to independent slices, thereby boosting MLP and performance. Moreover, as shown through our analysis and simulation, Freeway's policy can be implemented with a simple FIFO queue, which requires only minimum additional hardware over the baseline sOoO core. Our results show that Freeway is able to attain 12\% better performance than previous sOoO designs and delivers performance within 7\% of the MLP limits of the ideal sOoO execution.

Our analysis shows that Freeway is also more flexible than previous sOoO designs, as it requires one-sixth as many instruction queue entries to deliver similar performance and can tolerate 1.5x higher L1 cache latencies at similar performance. 
Further, while Freeway's MLP generation is complementary to that of an LLC prefetcher, we find that the Freeway generates significantly more MLP in general. 
However, our results also show that sOoO cores are less effective at larger instruction window sizes as MLP provides a smaller portion of the overall performance, with ILP becoming more important. Based on these results, and our data showing that Freeway captures the bulk of the available MLP, we conclude that future sOoO core research should focus on ILP, rather than capturing the limited MLP opportunities missed by Freeway.

\section{Acknowledgments}
This work was supported by the Knut and Alice Wallenberg Foundation through the Wallenberg Academy Fellows Program, the European Research Council (ERC) under the European Union's Horizon 2020 research and innovation program (grant No 715283), and the Research Council of Norway (NFR) grant 302279 to NTNU.

\bibliographystyle{ACM-Reference-Format}
\bibliography{ref}

\end{document}